\documentstyle[12pt,epsfig]{article}
\newlength{\dinwidth}
\newlength{\dinmargin}
\newcommand{\ba}{\begin{array}}
\newcommand{\ea}{\end{array}}
\newcommand{\beq}{\begin{equation}}
\newcommand{\eeq}{\end{equation}}
\newcommand{\bea}{\begin{eqnarray}}
\newcommand{\eea}{\end{eqnarray}}

\def\bce{\begin{center}}
\def\ece{\end{center}}

\def\nonu{\nonumber}

\def\pa{\partial}
\def\al{\alpha}

\def\de{\delta}
\def\De{\Delta}

\def\la{\lambda}

\def\si{\sigma}

\def\half{\frac{1}{2}}

\def\R{{\bf R}}

\def\S{{\bf S}}

\def\Kop{{\displaystyle \mathop{K}^{\circ}}{}}
\def\Gop{{\displaystyle \mathop{g}^{\circ}}{}}

\def\eps6{{\displaystyle \mathop{\epsilon}^{6}}{}}

\def\nab6{{\displaystyle \mathop{\nabla}^{6}}{}}

\def\to{\rightarrow}

\begin{document}
\thispagestyle{empty}
\addtocounter{page}{-1}
\begin{flushright}
%NSF-ITP 99-098\\
%SNUST 99-?\\
{\tt hep-th/0208137}\\
\end{flushright}
\vspace*{1.3cm}
\centerline{\Large \bf The 11-dimensional Metric for AdS/CFT RG Flows}
\vskip0.3cm
\centerline{\Large \bf with Common $SU(3)$ Invariance}
\vspace*{1.5cm} 
\centerline{{\bf Changhyun Ahn}$^1$ and {\bf Taichi Itoh}$^2$}
\vspace*{1.0cm}
\centerline{\it $^1$Department of Physics, 
Kyungpook National University, Taegu 702-701, Korea}
\vspace*{0.2cm}
\centerline{\it $^2$Department of Physics, 
Hanyang University, Seoul 133-791, Korea}
%\vskip0.3cm
%\vspace*{0.8cm}
%\centerline{\tt ahn@knu.ac.kr, \qquad taichi@hepth.hanyang.ac.kr}
\vskip2cm
\centerline{\bf Abstract}
\vspace*{0.5cm}

The compact 7-manifold arising in the compactification of 11-dimensional 
supergravity is described by the metric encoded in the vacuum expectation 
values(vevs) in $d=4$, ${\cal N}=8$ gauged supergravity. Especially, the 
space of $SU(3)$-singlet vevs contains various critical points and 
RG flows(domain walls) developing along $AdS_4$ radial coordinate.
Based on the nonlinear metric ansatz of de Wit-Nicolai-Warner, we show 
the geometric construction of the compact 7-manifold metric and find 
the local frames(siebenbeins) by decoding the $SU(3)$-singlet vevs 
into squashing and stretching parameters of the 7-manifold.     
Then the 11-dimensional metric for the whole $SU(3)$-invariant sector 
is obtained as a warped product of an asymptotically $AdS_4$ space with 
a squashed and stretched 7-sphere. We also discuss the difference 
in the 7-manifold between two sectors, namely 
$SU(3)\times U(1)$-invariant sector and $G_2$-invariant sector.
%The former has the common base manifold $\S^6$ as an Einstein-K\"{a}hler 
%3-fold ${\bf CP}^3$, whereas the latter has $\S^6$ as an homogeneous space 
%$G_2/SU(3)$. 
In spite of the difference in base 6-sphere, both sectors share the 
4-sphere of ${\bf CP}^2$ associated with the common $SU(3)$-invariance 
of various 7-manifolds. 

\vspace*{\fill}

\noindent
------------------\\
E-mail addresses: ahn@knu.ac.kr (C. Ahn), 
taichi@hepth.hanyang.ac.kr (T. Itoh)

%\centerline{Submitted to Nuclear Physics B}
\baselineskip=18pt
\newpage
\renewcommand{\theequation}{\arabic{section}\mbox{.}\arabic{equation}}

%%%%%%%%%%%%%%%%%%%%%%%%%%%%%%%%%%%%%%%%%%%%%%%%%%%%%%%%%%%%%%%%%%%%%%%%%%%%%
\section{Introduction}
\setcounter{equation}{0}
%%%%%%%%%%%%%%%%%%%%%%%%%%%%%%%%%%%%%%%%%%%%%%%%%%%%%%%%%%%%%%%%%%%%%%%%%%%%%

One of the crucial points of de Wit-Nicolai theory \cite{dn}
is the presence of warp factor $\Delta(x,y)$. 
When one reduces 11-dimensional
supergravity theory to four-dimensional ${\cal N}=8$ gauged supergravity, 
the four-dimensional
spacetime is warped by this factor which depends on both four-dimensional
coordinate $x^{\mu}$ and 7-dimensional internal coordinate $y^m$.   
This warp factor provides an understanding of the different relative scales
of the 11-dimensional solutions corresponding to 
the critical points in ${\cal N}=8$ gauged supergravity. 
One writes down 11-dimensional metric as warped product ansatz
\bea
ds_{11}^2 = 
ds_4^2 + ds_7^2 =
\frac{1}{\Delta(x,y)} \,g_{\mu \nu}(x) \,d x^{\mu} d x^{\nu}
+ G_{mn}(x,y) \,dy^m dy^n, 
\label{metric}
\eea 
where
$\mu, \nu=1, 2,  \cdots, 4$ and
$m,n=1, 2, \cdots, 7$.
The nonlinear metric ansatz in \cite{dnw} provides the explicit formula 
(\ref{7dmet}) for the 7-dimensional inverse metric $G^{mn}(x,y)$, which
is encoded by the warp factor $\Delta(x,y)$, 
28 Killing vectors $\Kop^{mIJ}(y)$ on 
the round 7-sphere(expressed through the Killing spinors) 
and 28-beins $u_{ij}^{~~IJ}(x), v_{ijIJ}(x)$
in four-dimensional gauged ${\cal N}=8$ supergravity: 
the dependence 
of $x$ appears in the first and last one.

An important aspect of the holographic
duals \cite{malda,witten,gubser} is the notion that the
radial coordinate of $AdS_4$ can be viewed as a measure of energy.
A supergravity kink description interpolating between 
$r \rightarrow \infty$ 
and $r \rightarrow -\infty$ can be interpreted as an explicit construction 
of the renormalization group(RG) flow
between the ultraviolet(UV) fixed point and 
the infrared(IR) fixed point of the three dimensional boundary 
field theory. To construct 
the superkink corresponding to the supergravity description
of the nonconformal RG flow, 
the most general four dimensional bulk metric which has 
a three-dimensional Poincare invariance takes the form 
\bea
g_{\mu \nu}(x) \,d x^{\mu} d x^{\nu} = e^{2A(r)} \,\eta_{\mu' \nu'}\,
d x^{\mu'} d x^{\nu'} + dr^2, 
\label{domain}
\eea
where 
$\eta_{\mu' \nu'} = (-, +, +)$,
$r=x^4$ is the coordinate transverse to the domain wall and 
the scale factor 
$A(r)$ behaves linearly in $r$ at UV and IR regions.

The 70 real scalars of ${\cal N}=8$ supergravity
parametrize \cite{cremmer} 
the coset space $E_{7(7)}/SU(8)$. The scalar potential
is a function of these 70 scalars and this number is too large to
be managed practically. From the possible embeddings of $SU(3)$,
the 70 scalars contain 6 singlets of $SU(3)$(three from 35 scalars
${\bf 35_v}$ plus three from 35 pseudo-scalars ${\bf 35_c}$). 
But the scalar potential depends on only four real fields due to the 
$SO(8)$-invariance of the potential and a larger invariance  of 
the $SU(3)$-invariant sector. The explicit construction of 
28-beins $u_{ij}^{~~IJ}(x)$ and $v_{ijIJ}(x)$ in terms of these fields
has been found in \cite{aw}.  
It is known \cite{warner} that there exist five nontrivial critical points
for the scalar potential of gauged ${\cal N}=8$ supergravity:
$SO(7)^{+}, SO(7)^{-}, G_2, SU(4)^{-}$ and $SU(3) \times U(1)$. 
Among them $G_2$-invariant 7-ellipsoid and $SU(3) \times U(1)$-invariant
stretched 7-ellipsoid are stable and supersymmetric.

By writing  de Wit-Nicolai potential in terms of 
a superpotential encoded in the 
T-tensor of a theory, the holographic
RG flow equations from ${\cal N}=8$, $SO(8)$-invariant UV fixed point 
to ${\cal N}=2$, $SU(3) \times U(1)$-invariant IR fixed point
were contructed in \cite{ap}.
Moreover,  
the holographic
RG flow equations from ${\cal N}=8$, $SO(8)$-invariant UV fixed point 
to
${\cal N}=1$, $G_2$-invariant IR 
fixed point were obtained in \cite{aw,ai}(See also \cite{ar2}). 
Contrary to $SU(3) \times U(1)$-invariant case, there exists an
algebraic relation in a complex eigenvalue of $A_1$ tensor
for the $G_2$-invariant case.
This observation was crucial for the correct minimization of the 
energy-functional.

The $M$-theory lift of a supersymmetric RG flow is achieved as follows.
First we impose the nontrivial $r$-dependence of vacuum expectation 
values(vevs) subject to the four-dimensional RG flow equations.
Then the geometric parameters in the 7-manifold metric at certain critical 
point are controlled by the RG flow equations so that they can be smoothly 
extrapolated from the critical point. Secondly we make an appropriate ansatz 
for the 11-dimensional three-form gauge field. If the ansatz is correct, 
the 11-dimensional Einstein-Maxwell(bosonic) equations can be finally solved 
by using the RG flow equations to complete the $M$-theory lift. 
Based on this prescription, an exact solution to the 11-dimensional 
bosonic equations corresponding to the $M$-theory lift of the ${\cal N}=2$, 
$SU(3) \times U(1)$-invariant RG flow was found in \cite{cpw}. 
Its K\"{a}hler structure was extensively studied in \cite{jlp}. 
Similarly, the $M$-theory lift of the ${\cal N}=1$, $G_2$-invariant 
RG flow was done in \cite{ai}.

In AdS/CFT context \cite{malda,witten,gubser}, the above two membrane flows 
are holographic dual of flows of the maximally supersymmetric ${\cal N}=8$ 
scalar-fermion theory in three-dimensions. It is still unclear how they are 
related both in the bulk supergravity and in the boundary field theory. 
In order to answer this question, we would like to achieve the $M$-theory 
lift of {\it whole} $SU(3)$-invariant sector, including the above five 
critical points and RG flows in $d=4$, ${\cal N}=8$ gauged supergravity. 
As the first step toward this goal, we need to know the complete metric 
(\ref{metric}) together with (\ref{domain}). To solve the 11-dimensional 
bosonic equations by utilizing the RG flow equations in \cite{aw}, 
we have to further make an appropriate ansatz for the 11-dimensional 
three-form gauge field. It will be a natural extension
of the Freund-Rubin parametrization \cite{fr} but will be more complicated 
due to its nontrivial $AdS_4$ radial coordinate dependence. Even though 
the nonlinear metric ansatz in \cite{dnw} provides the the 11-dimensional 
four-form gauge field strengths also,\footnote{
They are given by Eq.\ (7.6) in \cite{dn87}.} 
they are encoded in $SU(3)$-singlet vevs in a complicated way. 
In this paper, we concentrate on the 11-dimensional metric for the 
$SU(3)$-invariant sector and postpone the field strength ansatz 
as well as solving the 11-dimensional bosonic equations to future work.  

We begin our analysis in section 2 by summarizing relevant aspects 
of the 11-dimensional metric for ${\cal N}=2$, $SU(3)\times U(1)$-invariant 
membrane flow found in \cite{cpw}. 
In section 3, we review the 11-dimensional metric for 
${\cal N}=1$, $G_2$-invariant membrane flow found in \cite{ai}. 
In section 4, we summarize the four-dimensional RG flow equations \cite{aw}, 
namely the first-order BPS equations for the $SU(3)$-invariant sector 
in $d=4$, ${\cal N}=8$ gauged supergravity. 
In section 5, that is the main content of this paper, we show 
the geometric construction of the compact 7-manifold metric and find 
the local frames(siebenbeins) by decoding the $SU(3)$-singlet vevs 
into squashing and stretching parameters of the 7-manifold. 
We mainly use Hopf fibration on ${\bf CP}^3$ as global 
7-dimensional coordinates. 
In section 6, we summarize our results and will discuss about future 
direction. 
In appendix A, we list all the components of the 7-dimensional 
inverse metric generated from the metric formula (\ref{7dmet}) 
by using ${\bf R}^8$ embedding. 
In appendix B, some preliminaries of Hopf fibration are presented. 
In appendix C, we show the other set of global 7-dimensional coordinates 
useful to describe the $G_2$-invariant critical point. 

%%%%%%%%%%%%%%%%%%%%%%%%%%%%%%%%%%%%%%%%%%%%%%%%%%%%%%%%%%%%%%%%%%%%%%%%%%%%%
\section{The 11-dimensional metric for 
${\cal N}=2, SU(3)$$\times$$U(1)$-invariant membrane flow}
\setcounter{equation}{0}
%%%%%%%%%%%%%%%%%%%%%%%%%%%%%%%%%%%%%%%%%%%%%%%%%%%%%%%%%%%%%%%%%%%%%%%%%%%%%

The 28-beins $(u, v)$ or two vevs $(\rho, \chi)$ 
\footnote{The variables $\la$ and $\la'$ in \cite{ap} with 
$\al=0,\phi=\pi/2$ are related to 
$ \la=4\sqrt{2} \ln \rho, \la'=\sqrt{2} \chi$. In terms of 
$a,b,c$ and $d$ we will define later, $\rho =a^{1/4}=b^{-1/4}$ and 
$\chi =\cosh^{-1} c =\cosh^{-1} d$.}
are given by functions 
of the $AdS_4$ radial coordinate $r=x^4$. The metric formula  (\ref{7dmet})
generates the 7-dimensional metric from the two input data of $AdS_4$ vevs 
$(\rho,\chi)$. The $SU(3) \times U(1)$-invariant 
RG flow subject to the first order differential equations on two vevs 
\cite{ap} 
is a trajectory in $(\rho,\chi)$-plane and is parametrized by the 
$AdS_4$ radial coordinate. 

Let us introduce the standard metric 
corresponding to an ellipsoidally squashed
7-sphere with a stretched Hopf fiber.
The stretching factor is characterized by the vev $\chi$ and $\chi=0$ 
corresponds to the round 7-sphere. 
Then the isometry $SO(8)$ of 7-sphere breaks into $U(4)$.
The metric is also ellipsoidally squashed and it reduces the broken isometry
to $SU(3) \times U(1) \times U(1)$.
Let us introduce a diagonal $8 \times 8$ matrix 
\bea
Q_{AB}={\rm diag}\left(\rho(r)^{-2}, \dots, \rho(r)^{-2}, 
\rho(r)^{6}, \rho(r)^{6}\right).
\nonu
\eea
Recall that $A_1$ tensor of this theory has two distinct eigenvalues
with degeneracies 6, 2. This behavior
reflects also in the deformation matrix here. 
The metric on the deformed ${\bf R}^8$(Cartesian coordinates on 
${\bf R}^8$ are denoted by $X^A$ with $\sum_{A=1} (X^A)^2=1$) 
can be written as(See also 
\cite{jlp})
\bea
 ds^2(\rho(r), \chi(r))  
%& \equiv & 
%(dX, g dX) \equiv dX^A g_{AB} dX^B \nonu \\
  =  dX^A Q(r)^{-1}_{AB}\,dX^B 
%-\frac{1}{\xi(r,\mu)^2} \left( X^A \de_{AB} d X^B \right)^2 
+ \frac{\sinh^2 \chi(r)}{\xi(r,\mu)^2}\left(X^A J_{AB}\,dX^B\right)^2
\label{metricrhochi}
\eea
where
the quadratic form $\xi(r,\mu)^2 \equiv X^A Q_{AB}X^B$ 
is now given by using the 
parametrization of \cite{cpw}
\bea
\xi(r, \mu)^2 =\rho(r)^{-2} \cos^2 \mu + \rho(r)^6 \sin^2 \mu
\nonu
\eea
where $\mu$ is one of the 7-dimensional internal coordinates
and $\xi(r, \mu)^2$ becomes 1 when $\rho=1$.
For $\rho=1,\chi=0$, it provides the trivial vacuum of $SO(8)$ 
maximal supersymmetric critical point.
The antisymmetric K\"{a}hler form $J_{AB}$ has nonzero elements 
$J_{12}=J_{34}=J_{56}=J_{78}=1$. 

Applying the Killing vector together with 
the 28-beins $(u, v)$ to the metric formula (\ref{7dmet}), 
we obtain a ``raw'' 
inverse metric $\De(x,y)^{-1} G^{mn}(x,y)$ 
including the warp factor $\Delta(x,y)$ not yet determined. 
Substitution of this raw inverse metric into the definition
of warp factor 
\bea
\Delta(x,y)^{-1}\equiv\sqrt{\det(G^{mn}(x,y)\,\Gop_{\,np}(y))},
\label{defwarp}
\eea
where $\Gop_{\,np}(y)$ is a metric of the round 7-sphere, 
will provide a self-consistent equation for 
$\Delta(x,y)$. 
For the $SU(3) \times U(1)$-invariant RG flow, 
solving this equation gives rise to the warp factor
\bea
\Delta(r, \mu)=\left( \xi(r,\mu) \cosh \chi(r) \right)^{-{4 \over 3}}.
\label{warpsu3u1}
\eea
Then we substitute this warp factor 
into the ``raw'' inverse metric and obtain the 7-dimensional metric: 
\begin{equation}
ds_7^2 =G_{mn}(x, y) \,dy^m dy^n
=\sqrt{\Delta(r,\mu) } L^2 ds^2(\rho(r), \chi(r))
\label{7dmetric}
\end{equation}
together with (\ref{metricrhochi}) and (\ref{warpsu3u1}) 
where $L$ is a radius of round 7-sphere.
The metric (\ref{metricrhochi}) now is warped by a factor 
$\sqrt{\Delta(r,\mu)} $.
The nonlinear metric ansatz combines the 7-dimensional metric with
the four-dimensional metric with warp factor to yield the 11-dimensional 
warped metric characterized by  (\ref{metric}), (\ref{domain}), 
(\ref{7dmetric}), (\ref{warpsu3u1}) and (\ref{metricrhochi}).  

%%%%%%%%%%%%%%%%%%%%%%%%%%%%%%%%%%%%%%%%%%%%%%%%%%%%%%%%%%%%%%%%%%%%%%%%%%%%%
\section{The 11-dimensional metric for ${\cal N}=1,G_2$-invariant 
membrane flow}
\setcounter{equation}{0}
%%%%%%%%%%%%%%%%%%%%%%%%%%%%%%%%%%%%%%%%%%%%%%%%%%%%%%%%%%%%%%%%%%%%%%%%%%%%%

The two vevs $(\lambda, \alpha)$ 
\footnote{One can obtain these vevs from $SU(3)$-invariant sector by
restricting four arbitrary fields $\la, \la',\al, \phi$ to 
$\la'=\la$ and $\phi=\al$. This is equivalent to $c=a$ and $d=b$ where 
$a,b,c$ and $d$ are defined as the one in Section 5.}
are given by functions 
of the $AdS_4$ radial coordinate $r$. The metric formula (\ref{7dmet}) 
generates the 7-dimensional metric from the two input data of $AdS_4$ vevs 
$(\lambda,\alpha)$. The $G_2$-invariant RG flow subject to the first
order differential equations on these vevs \cite{aw,ai} 
is a trajectory in $(\lambda,\alpha)$-plane and is parametrized by the 
$AdS_4$ radial coordinate. 
We will use $(a,b)$ defined by
\begin{eqnarray}
a(r) &\equiv& \cosh\!\left(\frac{\lambda(r)}{\sqrt{2}}\right)
+\cos\alpha(r)\,\sinh\!\left(\frac{\lambda(r)}{\sqrt{2}}\right),\nonumber\\
b(r) &\equiv& \cosh\!\left(\frac{\lambda(r)}{\sqrt{2}}\right)
-\cos\alpha(r)\,\sinh\!\left(\frac{\lambda(r)}{\sqrt{2}}\right).
\label{abvevs}
\end{eqnarray}

Let us  introduce the standard metric of a 7-dimensional ellipsoid. 
Using the diagonal $8 \times 8$ matrix $Q_{AB}$ given by \footnote{
Note that the diagonal matrix here is different from the one in \cite{ai}
and is defined as the old one multiplied by $b(r)^2$.}
\bea
Q_{AB}={\rm diag}\left(b(r)^2, \dots, b(r)^2, a(r)^2\right),
\label{g2Q}
\eea
the metric of a 7-dimensional ellipsoid can be written as
\begin{equation}
ds_{EL(7)}^2(a(r), b(r)) =dX^A Q(r)^{-1}_{AB}\,dX^B. 
%-\frac{1}{\xi(r, \theta)^2}\left(X^A \delta_{AB}\,dX^B\right)^2. 
\label{7dmet2}
\end{equation}
%The standard metric (\ref{7dmet2}) 
This can be rewritten in terms of the 
7-dimensional coordinates $y^m$(See \cite{ai} for 
the explicit relations between ${\bf R}^8$ coordinates $X^A$ and $y^m$) 
such that
\begin{equation}
ds_{EL(7)}^2(a(r),b(r))=b(r)^{-2} \left[\,
a(r)^{-2}\,\xi(r, \theta)^2\,d\theta^2 +\sin^2 \theta\,d\Omega_6^2 
\,\right],
\label{7dmet3}
\end{equation}
where $\theta =y^7$ must be identified with the fifth coordinate 
in 11 dimensions and the quadratic form $\xi(r, \theta)^2$ is  given by
\bea
\xi(r, \theta)^2 =a(r)^2 \cos^2 \theta + b(r)^2 \sin^2 \theta
\label{xi}
\eea
which becomes 1 for $a=b=1$ corresponding to $SO(8)$ trivial vacuum.

Note that 
the geometric parameters $(a,b)$ for the 7-ellipsoid can be identified 
with the two vevs $(a,b)$ defined in (\ref{abvevs}). 
This was the reason why we prefer the vevs $(a,b)$ rather than 
$(\la, \al)$.
Applying the Killing vector together with 
the 28-beins to the metric formula (\ref{7dmet}) like we did before, 
we obtain a ``raw'' 
inverse metric $\Delta(x,y)^{-1}G^{mn}(x,y)$ 
including the warp factor $\Delta(x,y)$. 
For the $G_2$-invariant RG flow, solving the condition yields the warp factor
\bea
\Delta(r, \theta)=a(r)^{-1}\,\xi(r,\theta)^{-{4 \over 3}}.
\label{warpg2}
\eea
Then we substitute this warp factor 
into the raw inverse metric  and obtain the 7-dimensional metric as follows: 
\begin{equation}
ds_7^2 =G_{mn}(x, y) \,dy^m dy^n
=\sqrt{\Delta(r,\theta)\,a(r)}\;b(r)^2 L^2 
ds_{EL(7)}^2(a(r), b(r)).
\label{ellip}
\end{equation}
The 7-dimensional metric (\ref{7dmet2}) is warped by a factor 
$\sqrt{\Delta(r,\theta)\,a(r)}$.
The nonlinear metric ansatz with the warp factor yields the warped 
11-dimensional metric described by (\ref{metric}), (\ref{domain}), 
(\ref{ellip}), (\ref{7dmet2}) and (\ref{warpg2}).  

%%%%%%%%%%%%%%%%%%%%%%%%%%%%%%%%%%%%%%%%%%%%%%%%%%%%%%%%%%%%%%%%%%%%%%%%%%%%%
\section{ Holographic RG flow for $SU(3)$-invariant sector}
\setcounter{equation}{0}
%%%%%%%%%%%%%%%%%%%%%%%%%%%%%%%%%%%%%%%%%%%%%%%%%%%%%%%%%%%%%%%%%%%%%%%%%%%%%

It is known \cite{warner} that 
$SU(3)$-singlet space with a breaking of the 
$SO(8)$ gauge group into a group which contains $SU(3)$
may be written as four real parameters 
$\la(r), \la'(r),\al(r)$ and $\phi(r)$. 
The vacuum expectation value of 56-bein for the 
$SU(3)$-singlet space, that is an invariant subspace 
under a particular $SU(3)$ subgroup of $SO(8)$, 
can be parametrized by
\begin{eqnarray}
\phi_{ijkl} & = &  \; \la(r) \; \mbox{cos}\alpha(r)\; Y^{1\;+}_{ijkl} +
\la(r) \; \mbox{sin}\alpha(r) \; Y^{1\;-}_{ijkl} + \la'(r) 
\; \mbox{cos} \phi(r) \;
%\mbox{cos} (\theta + \psi) 
\; Y^{2\;+}_{ijkl} 
 + \la'(r) \; \mbox{sin} \phi(r) \; 
%\mbox{cos}(\theta - \psi) \; 
Y^{2\;-}_{ijkl} \; ,
\nonu
\end{eqnarray}
where the scalar and pseudo-scalar singlets of $SU(3)$ are given by 
\begin{eqnarray}
  Y^{1\;\pm}_{ijkl} &=& \varepsilon_{\pm} \left[ \; (\de^{1234}_{ijkl} \pm 
\de^{5678}_{ijkl})+
 (\de^{1256}_{ijkl} \pm \de^{3478}_{ijkl})+(\de^{3456}_{ijkl}
 \pm \de^{1278}_{ijkl}) \; \right],
\nonu \\
       Y^{2\;\pm}_{ijkl} &=& \varepsilon_{\pm} \left[ -(\de^{1357}_{ijkl}
\pm \de^{2468}_{ijkl})+(\de^{2457}_{ijkl} \pm
\de^{1368}_{ijkl})+(\de^{2367}_{ijkl} \pm \de^{1458}_{ijkl}) +
 (\de^{1467}_{ijkl} \pm \de^{2358}_{ijkl}) \right]. 
\nonu
\end{eqnarray} 
Here  $\varepsilon_{+}=1$ and $\varepsilon_{-}=i$ and $+$ gives the scalars
and $-$ gives the pseudo-scalars of ${\cal N}=8$ supergravity. 
The four scalars $\la(r), \la'(r), \al(r)$ and $\phi(r)$ 
in the $SU(3)$-singlet vevs 
parametrize an $SU(3)$-invariant subspace of the complete
scalar manifold $E_{7(7)}/SU(8)$.
The 56-bein ${\cal V}(x)$ preserving the $SU(3)$-singlet space  
is a $56 \times 56$ matrix
whose elements are some functions of four fields $\la(r), \la'(r), 
\al(r)$ and $\phi(r)$ obtained 
by exponentiating the above vacuum expectation value $\phi_{ijkl}$.
Then the 28-beins, $u$ and $v$ in terms of these fields, can be obtained as 
the $28 \times 28$ matrices given in the appendix
A of \cite{aw}.

It turned out \cite{aw} that 
$A_1$ tensor has three distinct complex
eigenvalues, $z_1$, $z_2$ and $z_3$ 
with degeneracies 6, 1, and 1 respectively and has the following form
\bea
A_1^{\;IJ} =\mbox{diag} \left(z_1, z_1, z_1, z_1, z_1,
z_1, z_2, z_3 \right), \nonu
\eea
where the eigenvalues are some 
functions of $\la(r), \la'(r), \al(r)$ and $\phi(r)$.
In particular,
\begin{eqnarray}
z_3(\la,\la',\al,\phi) & = & 6e^{i(\alpha + 2\phi)}p^2qr^2t^2
 + 6e^{2i(\alpha + \phi)}pq^2r^2t^2 + p^3 \left(r^4 + e^{4i\phi}t^4 \right)  +
e^{3i\alpha}q^3 \left(r^4 + e^{4i\phi}t^4 \right),
\nonu
\end{eqnarray}
and we denote hyperbolic functions of $\la(r)$ and $\la'(r)$ by the following 
quantities for simplicity
\bea 
& & p \equiv \cosh \left(\frac{\la(r)}{2\sqrt{2}}\right), \;\; q \equiv
\sinh\left(\frac{\la(r)}{2\sqrt{2}}\right), 
\;\; r \equiv \cosh\left(\frac{\la'(r)}{2\sqrt{2}}\right), \;\; t
\equiv \sinh\left(\frac{\la'(r)}{2\sqrt{2}}\right). 
\nonu
\eea 
We refer to \cite{aw} for explicit expressions of $z_1$ and $z_2$.

The superpotential is one of the eigenvalues of $A_1$ tensor and
the supergravity potential\footnote{
The scalar potential 
can be written, by combining all the components of
$A_1, A_2$ tensors,  as
\bea 
 V(\la, \la', \al, \phi)  
& = & \frac{1}{2} g^2 \left( s'^4 \left[ (x^2+3)c^3 + 4x^2v^3s^3 -
3v(x^2-1)s^3 + 12xv^2cs^2 - 6(x-1)cs^2 + 6(x+1)c^2sv \right] \right.  \nonu \\ 
&& \left. + 2s'^2 \left[ 
2(c^3+v^3s^3) + 3(x+1)vs^3 + 6xv^2cs^2 - 3(x-1)cs^2 -
6c \right]- 12c \right), 
\nonu 
\eea 
where we introduce
the following quantities:
$ 
c \equiv \cosh\left(\frac{\la}{\sqrt{2}}\right)$,  
$s \equiv \sinh\left(\frac{\la}{\sqrt{2}}\right)$,  
$c' \equiv \cosh\left(\frac{\la'}{\sqrt{2}}\right)$, 
$s' \equiv \sinh\left(\frac{\la'}{\sqrt{2}}\right)$, 
$v \equiv \cos\alpha$, 
and $x \equiv \cos2\phi$.   
} 
can be written in terms of superpotential as follows 
\bea
W(\la, \la', \al, \phi)  & = &  |z_3|, \nonu \\
V(\la, \la',\al, \phi) & = & g^2 \left[  \frac{16}{3} \left(\partial_{\la}
W \right)^2  + \frac{2}{3p^2 q^2}
\left(\partial_{\al}
W \right)^2 + 4 
\left(\partial_{\la'} 
W  \right)^2+ \frac{1}{2r^2 t^2} \left(\partial_{\phi} 
W \right)^2 - 6  W^2 \right].
\label{potandsuper}
\eea
The flow equations \cite{aw} we are interested in are
\bea  
\partial_{r}\la(r) & = & 
- \frac{8\sqrt{2}}{3}\; g \;\partial_{\la} W(\la,\la',\al,\phi) ,\nonu
\\
\partial_{r}\la'(r) & = & 
- 2 \sqrt{2} \; g \; \pa_{\la'} W(\la,\la',\al,\phi), \nonu \\
\partial_{r}\alpha(r) & = & -
\frac{\sqrt{2}}{3p^2 q^2} \; g \; \pa_{\al} W(\la,\la',\al,\phi), \nonu \\ 
\partial_{r}\phi(r) & = & 
- \frac{\sqrt{2}}{4r^2 t^2} \; g \; \pa_{\phi} W(\la,\la',\al,\phi) , \nonu \\ 
\partial_{r}A(r) & = &  \sqrt{2} \;g \;  W(\la,\la',\al,\phi).
\label{first}
\eea
There exist two supersymmetric critical points of
both a scalar potential  and a superpotential: ${\cal N}=1$ supersymmetric
critical point with $G_2$-symmetry and ${\cal N}=2$ supersymmetric one
with $SU(3) \times U(1)$-symmetry. Also there are three nonsupersymmetric
critical points with $SO(7)^{+}, SO(7)^{-}$ and $SU(4)^{-}$-symmetries.

%%%%%%%%%%%%%%%%%%%%%%%%%%%%%%%%%%%%%%%%%%%%%%%%%%%%%%%%%%%%%%%%%%%%%%%%%%%%%
\section{The 11-dimensional metric for $AdS_4$ RG flows with common 
$SU(3)$ invariance}
%The $M$-theory lift of the $SU(3)$-invariant holographic RG flow}
\setcounter{equation}{0}
\setcounter{subsection}{0}
%%%%%%%%%%%%%%%%%%%%%%%%%%%%%%%%%%%%%%%%%%%%%%%%%%%%%%%%%%%%%%%%%%%%%%%%%%%%%

In this section, we will give an ansatz for the generic $SU(3)$-invariant 
metric in terms of squashing deformation and the K\"{a}hler form $J$ 
by using the $\R^8$ vector $X=R\,x$. Then we invert the metric and change the 
$\R^8$ basis from $X$ to $x$ to compare the ansatz with the inverse 
metric generated by the de Wit-Nicolai-Warner formula \cite{dnw,dn87}. 

%%%%%%%%%%%%%%%%%%%%%%%%%%%%%%%%%%%%%%%%%%%%%%%%%%%%%%%%%%%%%%%%%%%%%%%%%%%%%
\subsection{\bf The compact 7-manifold metric encoded in data of 
$d=4$, ${\cal N}=8$ gauged supergravity}
%%%%%%%%%%%%%%%%%%%%%%%%%%%%%%%%%%%%%%%%%%%%%%%%%%%%%%%%%%%%%%%%%%%%%%%%%%%%%

The consistency under the Kaluza-Klein compactification of 11-dimensional 
supergravity, or $M$-theory, requires that the 11-dimensional metric for 
RG flows with common $SU(3)$ invariance is {\it not} simply a metric of 
product space {\it but} Eq.\ (\ref{metric}) for the warped product of 
an asymptotically $AdS_4$ space, or a domain wall, with a compact 
7-dimensional manifold. 
Moreover, the 7-dimensional space becomes a warped, squashed and stretched 
${\bf S}^7$ and its metric is uniquely determined through the nonlinear 
metric ansatz developed in \cite{dnw,dn87}. 
The warped 7-dimensional inverse metric is given by
\begin{equation}
G^{mn}(x,y)=\frac{1}{2}\,\Delta(x,y)
\left[\Kop^{mIJ}\Kop^{nKL}+(m \leftrightarrow n)\right]
\left(u_{ij}^{~~IJ}(x)+v_{ijIJ}(x)\right)
\left(\overline{u}^{\,ij}_{~~KL}(x)+\overline{v}^{\,ijKL}(x)\right)
\label{7dmet}
\end{equation}
where $\Kop^{mIJ}$ denotes the Killing vector on the round ${\bf S}^7$ 
with 7-dimensional coordinate indices $m, n=5,\dots,11$ as well as $SO(8)$ 
vector indices $I, J=1,\dots,8$. The $u_{ij}^{~~IJ}$ and $v_{ijIJ}$ are  
28-beins in 4-dimensional gauged supergravity and are parametrized 
by the $AdS_4$ vacuum expectation values(vevs), $\lambda$, $\lambda^{'}$, 
$\alpha$ and $\phi$, associated with the spontaneous compactification 
of 11-dimensional supergravity. 

The 28-beins $(u, v)$ or four vevs $(\lambda, \la', \alpha, \phi)$ 
are given by functions 
of the $AdS_4$ radial coordinate $r=x^4$. The metric formula (\ref{7dmet}) 
generates the 7-dimensional metric from the four input data of $AdS_4$ vevs 
$(\lambda, \la', \alpha, \phi)$. The RG flow subject to Eq.\ (\ref{first}) 
is a trajectory in the $SU(3)$-singlet space spanned by 
$(\lambda, \la', \alpha, \phi)$ and is parametrized by the 
$AdS_4$ radial coordinate. 
Hereafter, instead of $(\lambda, \la', \alpha, \phi)$, 
we will use $(a, b, c, d)$ defined by
\begin{eqnarray}
a(r) &\equiv& \cosh\!\left(\frac{\lambda(r)}{\sqrt{2}}\right)
+\cos\alpha(r)\,\sinh\!\left(\frac{\lambda(r)}{\sqrt{2}}\right),\nonu \\
%\;\;
b(r) &\equiv& \cosh\!\left(\frac{\lambda(r)}{\sqrt{2}}\right)
-\cos\alpha(r)\,\sinh\!\left(\frac{\lambda(r)}{\sqrt{2}}\right), \nonu \\
c(r) &\equiv& \cosh\!\left(\frac{\lambda'(r)}{\sqrt{2}}\right)
+\cos\phi(r) \,\sinh\!\left(\frac{\lambda'(r)}{\sqrt{2}}\right),\nonu \\
%\;\;
d(r) &\equiv& \cosh\!\left(\frac{\lambda'(r)}{\sqrt{2}}\right)
-\cos\phi(r) \,\sinh\!\left(\frac{\lambda'(r)}{\sqrt{2}}\right).
\nonu
\end{eqnarray}

The inverse metric for the 7-manifold generated by the 
de Wit-Nicolai-Warner(dWNW) formula (\ref{7dmet}) is encoded in the data 
of 28-beins, namely $a$, $b$, $c$, $d$ given above. Therefore, to get the 
metric available for practical purposes, we have to decode the 28-beins 
into the deformation parameters of the compact 7-manifold. This can be done 
by comparing the encoded inverse metric with the inverse metric given 
by some appropriate ansatz. Since we do not know what are the proper 
coordinates describing the compact 7-manifold yet, we will consider the 
$\R^8$ embedding by using the Cartesian coordinates on $\R^8$, namely 
$x^A, A=1, \cdots, 8$. Now the Killing vectors defined on the round 
$\S^7$ with radius $L$ are given by
$$
\Kop_A^{~~IJ}=L\,(\Gamma^{IJ})_{BC}(x^B \pa_A x^C -x^C \pa_A x^B)
=L\left[\,x^B (\Gamma^{IJ})_{BA} -x^C (\Gamma^{IJ})_{AC}\right]
$$
where the $\R^8$ coordinates $x^A$'s are constrained on the unit round $\S^7$, 
$\sum_{A=1}^8 ( x^A )^2=1$, and $\Gamma^{IJ}$ are the $SO(8)$ generators 
given in \cite{aw2,ai}. 
Substituting this into the formula (\ref{7dmet}) generates 
the inverse metric divided by the warp factor, namely 
$\Delta^{-1}G^{AB}$, described by the $\R^8$ coordinates $x^A$'s. 
We list all of its components in appendix A. 
Our goal is to reproduce the same inverse metric via purely geometric 
construction and finally to determine the metric and the warp factor
separately. 

The complication coming from using the $\R^8$ coordinates is that 
the K\"{a}hler form $J$ defined by $J^2 =-I$ is not standard form in 
the coordinates $x^A$'s. By taking the $SU(4)^-$-invariant limit 
$a=b=1$, $d=c$ \cite{pw} in the raw inverse metric in appendix A, 
we obtain
$$
\Delta^{-1}G^{AB} = c^2 (\delta^{AB} -x^A x^B) +(1-c^2)
\,(\tilde{J}_{AC}\,x^C)(\tilde{J}_{BD}\,x^D)
$$
from which one can read the K\"{a}hler form $\tilde{J}$ as 
an $8\times8$ matrix given by Eq.\ (\ref{kahler0}) in appendix A.
Therefore to get the standard K\"{a}hler form $J$ given by
\bea
J_{12}=J_{34}=J_{56}=J_{78}=1,
\label{kahler}
\eea
we must transform the $\R^8$ vector $x$ to another one $X=R\,x$ by using
the $8\times8$ orthogonal matrix $R$ given by Eq.\ (\ref{RJR}) in appendix A.

%%%%%%%%%%%%%%%%%%%%%%%%%%%%%%%%%%%%%%%%%%%%%%%%%%%%%%%%%%%%%%%%%%%%%%%%%%%%%
\subsection{\bf Geometric construction of the compact 7-manifold metric}
%%%%%%%%%%%%%%%%%%%%%%%%%%%%%%%%%%%%%%%%%%%%%%%%%%%%%%%%%%%%%%%%%%%%%%%%%%%%%

Geometric implication of ${\cal N}=2$, $SU(3)\times U(1)$-invariant 
metric is as follows.
First, turning off the ellipsoidal deformation, the metric of compact 
7-manifold is given by Hopf fibration on ${\bf CP}^3$ with a stretched 
Hopf fiber. Fubini-Study metric on ${\bf CP}^3$ has $SU(4)$ invariance 
as can be seen from ${\bf CP}^3 \equiv SU(4)/(SU(3)\times U(1))$.
Then turning on the ellipsoidal deformation breaks the $SU(4)$ invariance 
of ${\bf CP}^3$ down to its $SU(3)\times U(1)$ subgroup preserving 
the $SU(3)$ invariance of Fubini-Study metric on 
${\bf CP}^2 \subset {\bf CP}^3$ as well as the $U(1)$ symmetry along 
the stretched Hopf fiber. Therefore to get the $SU(3)$-invariant metric, 
one can further break the $U(1)$ symmetry as long as the Fubini-Study 
metric on ${\bf CP}^2$ is preserved. Since the $\S^5$ given by Hopf 
fibration on ${\bf CP}^2$ is embedded in $\R^6$ spanned by 
$X^1,\cdots,X^6$, the first six diagonal components of 
the deformation matrix $Q$ must be the same. This is the same as 
in $SU(3)\times U(1)$-invariant case. Recall that the $\S^1$ of $U(1)$ 
Hopf fiber on ${\bf CP}^3$ is embedded in $\R^2$ spanned by $X^7$, $X^8$.
Now one can break this $U(1)$ symmetry by choosing different scaling factors 
in the last two diagonal components of $Q$. One can also decompose the Hopf 
fiber $(X,J dX)\equiv X^A J_{AB}\,dX^B$ into $SU(3)$-invariant pieces.  

Thus we are led to the ansatz for the $SU(3)$-invariant unwarped metric:
\beq
ds_0^2 = (dX, Q^{-1}dX)+ \frac{\gamma}{\xi^2}
\Bigl[(U, J dU)+ \zeta_1 (V_1, J dV_2) +\zeta_2 (V_2, J dV_1)\Bigr]^2 
\label{su3ans}
\eeq
where the deformation matrix $Q$ is given by
$$
Q = {\rm diag}(\eta,\cdots,\eta,\eta_1,\eta_2),
$$
so that $\xi^2 \equiv (X, Q X)$ can be the $SU(3)$-invariant norm on 
the 7-sphere. The $\R^8$ coordinates $X$ are restricted on the round 
$\S^7$, $\sum_{A=1}^8 (X^A)^2 =1$, and subject to the $SU(3)$-invariant 
decomposition $X=U+V_1+V_2$ with
$$
U = (X^1,\cdots, X^6, 0, 0), \quad
V_1 = (0,\cdots, 0, X^7, 0), \quad
V_2 = (0,\cdots, 0, 0, X^8).
$$
By introducing two more parameters $\zeta_1$, $\zeta_2$, the Hopf fiber 
$(X,J dX)$ is decomposed into the $SU(3)$-invariant pieces $(U,JdU)$, 
$(V_1,JdV_2)$, $(V_2,JdV_1)$. Now the $U(1)$ symmetry of 
$(X,J dX)$ is fully broken unless $\zeta_1 =\zeta_2=1$. 

From the ansatz (\ref{su3ans}), one can read the matrix $g$ as follows.
\beq
g = Q^{-1} -\frac{1}{\xi^2}\,X X^T + \frac{\gamma}{\xi^2}\,F
\label{gmet}
\eeq
where we have defined $F$ as
\beq
F = (J U+ \zeta_1 J V_1 +\zeta_2 J V_2)\,
(J U+ \zeta_1 J V_1 +\zeta_2 J V_2)^T. 
\label{Ftens}
\eeq
We notice that the first two terms in $g$ combine into the projection 
operator for the direction transverse to $QX$. This must be the case 
for the whole of $g$ since the metric $g$ describes the embedding of 
the 7-sphere into $\R^8$ by restricting the $\R^8$ into the subspace 
transverse to $QX$. Thus we require that $F$ must project out the 
deformed vector $QX$, namely $FQX = 0$ which restricts $\zeta_1$, 
$\zeta_2$ to be $\zeta_2 \eta_1 =\zeta_1 \eta_2$ and makes $(FQ)^2$ 
become proportional to $FQ$. We further require that
\beq
 (FQ)^2 = \xi^2 FQ, \label{FQ2}
\eeq
which can be achieved by choosing
$$
\zeta_1=\frac{1}{\zeta_2}=\sqrt{\hspace{1mm}\frac{\eta_1}{\eta_2}}.
$$

The deformation parameters $\eta$, $\eta_1$, $\eta_2$, $\gamma$ 
must be determined by comparing the inverse of $g$ with the inverse metric 
generated by dWNW formula. However, the $8\times 8$ matrix $g$ is a 
projection operator describing the $\R^8$ embedding of the compact 7-manifold 
and is rank 7. It does not have its inverse in ordinary sense. 
Therefore we have to define the inverse of $g$, say $g^{-1}$, 
as an $8\times8$ matrix satisfying
\beq
g^{-1} g\,g^{-1} = g^{-1} \;\longleftrightarrow\; g\,g^{-1} g = g.
\label{inversion}
\eeq
By looking at the $SU(3)$-invariant norm $\xi^2 =(X, Q X)$, we realize 
that $QX$ is a vector dual to $X$. This implies that $g^{-1}$ must be 
a projection operator transverse to $X$ as $g$ is so for $QX$. 
Now we suppose that $g^{-1}$ is given by
\beq
 g^{-1} = Q -\frac{1}{\xi^2}\,(QX) (QX)^T 
-\left(\frac{\gamma}{\gamma+1}\right)\frac{1}{\xi^2}\,QFQ.
\label{gbmet}
\eeq
which in fact projects out $X$ by using $FQX=0$. One can see by using 
Eq.\ (\ref{FQ2}) that $g$, $g^{-1}$ in Eqs.\ (\ref{gmet}), (\ref{gbmet})
indeed satisfy the inversion relation (\ref{inversion}). 

Thus the inverse metric $g^{-1}$ in Eq.\ (\ref{gbmet}) can be 
compared with the warped inverse metric $G^{-1}$ obtained by dWNW formula. 
They are related via
\beq
 \Delta^{-1} G^{-1} = \xi^2 N R^T g^{-1} R, 
\label{deltag}
\eeq
where $\xi^2$ in the right hand side is necessary to get the left hand side 
just as a polynomial of the $\R^8$ coordinates $x^A$'s as in appendix A. 
The normalization factor $N$, as well as the deformation 
parameters $\eta$, $\eta_1$, $\eta_2$ and $\gamma$, will be determined 
as a function of $a$, $b$, $c$, $d$. We also have to replace 
the $\R^8$ vector $X$ with $R\,x$ in the right hand side. 
Substituting the results in appendix A into the left hand side, 
one can get
$$
\eta_1 = \frac{ac}{bd}\,\eta, \quad
\eta_2 = \frac{ad}{bc}\,\eta, \quad
N = \frac{bcd}{\eta^2}, \quad
\gamma = \frac{cd}{ab} - 1. 
$$
where $\eta$ can be fixed by writing it 
like as $\eta =a^{m_1} b^{m_2} c^{m_3} d^{m_4}$ and using the
two limit values: $SU(3) \times U(1)$-invariant case and $G_2$-invariant case.
For the former, we have $b=1/a=\rho^{-4}$ and $d=c=\cosh \chi$. In this case,
$\eta$ was given as $\rho^{-2}$. Therefore we should have $m_3=-m_4$ and 
$m_1-m_2=-1/2$. 
For the latter, one has $c=a$ and $d=b$ and $\eta$ was given as $b^2$. 
Therefore $m_1+m_3=0$ and $m_2+m_4=2$. The exponents of $m_1$, $m_2$, $m_3$ 
and $m_4$ are uniquely determined to yield
\beq
\eta   =\frac{a^{3\over 4} b^{5\over 4} d^{3\over 4}}{c^{3\over 4}},\quad
\eta_1 =\frac{a^{7\over 4} b^{1\over 4} c^{1\over 4}}{d^{1\over 4}},\quad
\eta_2 =\frac{a^{7\over 4} b^{1\over 4} d^{7\over 4}}{c^{7\over 4}},\quad  
N =\frac{c^{5\over 2}}{a^{3\over 2} b^{3\over 2} d^{1\over 2}}L^{-2}.
\label{parame}
\eeq

Here a comment on $G_2$-invariant limit must be in order. In the limit, 
we have $c=a$ and $d=b$ so that Eq.\ (\ref{parame}) 
provides $\eta=\eta_2 =b^2$ and $\eta_1=a^2$. Hence we get 
$Q={\rm diag\,}(b^2,\dots,b^2,a^2,b^2)$ which is different from $Q$ 
in section 3 (See Eq.\ (\ref{g2Q})) by the interchange between the seventh 
and the eighth components. To fix this mismatch, we only have to 
interchange $X^7$ with $X^8$ by an orthogonal transformation. However, 
if we do so the K\"{a}hler form $J$ is also transformed and is no longer 
the standard form given in Eq.\ (\ref{kahler}). Nevertheless, this does 
not cause any problem in $G_2$-invariant case as well as in 
$SO(7)^{\pm}$-invariant cases since the 7-dimensional metric has no 
contribution from the K\"{a}hler form $J$.

%%%%%%%%%%%%%%%%%%%%%%%%%%%%%%%%%%%%%%%%%%%%%%%%%%%%%%%%%%%%%%%%%%%%%%%%%%%%%
\subsection{\bf The local frames for the compact 7-manifold}
%%%%%%%%%%%%%%%%%%%%%%%%%%%%%%%%%%%%%%%%%%%%%%%%%%%%%%%%%%%%%%%%%%%%%%%%%%%%%

Now that we have decoded the encoded output of dWNW formula given 
in appendix A by determining all the geometric parameters in the 
$SU(3)$-invariant ansatz (\ref{su3ans}). 
They are given by Eq.\ (\ref{parame}). 
However, the dWNW formula generates the involved inverse metric 
$\Delta^{-1}G^{-1}$ without giving the warp factor $\Delta$ separately. 
Therefore to get the full 7-dimensional metric, we have to separate out 
the warp factor from the obtained results. Recall that the inverse metric 
$g^{-1}$ is related to $\Delta^{-1}G^{-1}$ via Eq.\ (\ref{deltag}).
Inverting Eq.\ (\ref{deltag}) provides the warped 7-dimensional metric $G$ 
involving the warp factor. Since $g^{-1}$ is the inverse of $g$ in 
Eq.\ (\ref{gmet}), $G$ is given in the $\R^8$ basis $X$ by
\beq
G = \frac{1}{\Delta\,\xi^2 N}\,g \equiv
\frac{L^2}{\xi^2 \Delta}
\left(\frac{a^{3\over2}b^{3\over2}d^{1\over2}}{c^{5\over2}}\right)
\left[\,Q^{-1} -\frac{1}{\xi^2}\,X X^T + \frac{\gamma}{\xi^2}\,F\,\right].
\label{Gfull}
\eeq

To determine the warp factor $\Delta$, the easiest way is to find out 
the 7-dimensional local frames (or siebenbeins) $E^{\,i}$ ($i=1,\dots,7)$ 
defined as
\bea
ds_7^2 \equiv (dX, G\,dX) = L^2 \sum_{i=1}^7 E^{\,i} \otimes E^{\,i}.
\nonumber
\eea
By using the wedge product of $E^{\,i}$'s, the defining equation 
(\ref{defwarp}) of the warp factor $\Delta$ can be written as
\beq
\Delta \equiv (\Omega_7)^{-1}\bigwedge_{i=1}^7 E^{\,i},
\label{warp}
\eeq
where $\Omega_7$ is the volume element of the unit round $\S^7$. 
For convenience, let us introduce the unwarped local frames $e^{\,i}$ 
$(i=1,\cdots,7)$ defined by
\beq
ds_0^2 \equiv (dX, g\,dX) = \sum_{i=1}^7 e^{\,i} \otimes e^{\,i},
\label{nline}
\eeq
factorizing the $\Delta$ dependence of $E^{\,i}$'s. 
Eq.\ (\ref{warp}) then turns out to be 
the self-consistent equation for $\Delta$:
\beq
\Delta = \left[\frac{1}{\xi \Delta^{1\over2}}
\left(\frac{a^{3\over4}b^{3\over4}d^{1\over4}}{c^{5\over4}}\right)\right]^7 
(\Omega_7)^{-1} \bigwedge_{i=1}^7 e^{\,i}
\label{self}
\eeq
where the wedge product of $e^{\,i}$'s is calculable without knowing the 
warp factor. Hence in order to determine the warp factor $\Delta$, 
we only have to find out the unwarped frames $e^{\,i}$'s and 
to calculate the wedge product of them.

In terms of deformation parameters and the K\"{a}hler form $J$, 
the line element $ds_0^2$ in Eq.\ (\ref{nline}) is written 
explicitly as (See Eq.\ (\ref{su3ans}))
\bea
ds_0^2 &=& \frac{1}{\eta}\,(dU)^2 +\frac{1}{\eta_1}\,(dV_1)^2 
+\frac{1}{\eta_2}\,(dV_2)^2 \nonumber\\ &&
+\frac{\gamma}{\xi^2} \Biggl[(U,JdU)
+\sqrt{\hspace{1mm}\frac{\eta_1}{\eta_2}}\,(V_1,JdV_2)
+\sqrt{\hspace{1mm}\frac{\eta_2}{\eta_1}} \,(V_2,JdV_1)\Biggr]^2,
\label{dbs}
\eea
with $\gamma=cd/ab-1$. We will use the same 7-dimensional coordinatization 
of $U$, $V_1$, $V_2$ as in \cite{cpw}, that is
$$
U = u \cos \mu, \quad V_1 = (0,\cdots,0,\cos \psi \sin \mu,0), 
\quad V_2 =(0,\cdots,0,0,\sin \psi \sin \mu),
$$
where $u=(u^1,\dots,u^6,0,0)$ subject to the constraint $(u,u)=1$ 
describing the unit round $\S^5$. 
The deformed norm $\xi^2 \equiv (X,QX)$ then becomes
\beq
\xi^2 = \eta\cos^2\!\mu 
+(\eta_1 \cos^2\!\psi +\eta_2 \sin^2\!\psi)\sin^2\!\mu
\label{dnorm}
\eeq
which becomes 1 in the $SO(8)$-invariant limit $\eta=\eta_1=\eta_2=1$ 
to ensure the correct normalization of the unwarped metric $g$. 
The vector $u$ spans the $\S^5$ given by Hopf fibration on ${\bf CP}^2$ base.
This can be understood by rewriting $(du)^2$ as
\bea
(du)^2 = (du)^2-(u,Jdu)^2 +(u,Jdu)^2 \equiv ds_{FS(2)}^2 +(u,Jdu)^2,
\nonumber
\eea
where $ds_{FS(2)}^2 \equiv (du)^2-(u,Jdu)^2$ denotes the Fubini-Study 
metric on ${\bf CP}^2 \cong\S^4$ and $(u,J du)$ is the Hopf fiber on it.
As mentioned before, the $SU(3)$-invariant deformation must preserve 
at least the Fubini-Study metric on ${\bf CP}^2$. The $U(1)$ symmetry 
associated with the Hopf fiber $(u,Jdu)$ is a maximal circle of 
$\S^5\subset{\bf CP}^3$ and is always preserved. The $U(1)$ symmetry 
preserved in the $SU(3)\times U(1)$-invariant sector 
but broken in other cases is another $U(1)$ symmetry related to $(X,JdX)$, 
namely the Hopf fiber on ${\bf CP}^3$.

In $SU(3)\times U(1)$ limit, one of the local frames must be fixed 
to the direction of the Hopf fiber $(X,JdX)$ \cite{cpw}. 
Therefore it seems plausible that one of the local frames for the 
$SU(3)$-invariant 7-manifold, say $e^{7}$, is given by
\beq
e^{7} =\xi^{-1}\sqrt{\mathstrut{\gamma}+1}\left[(U,JdU)
+\sqrt{\hspace{1mm}\frac{\eta_1}{\eta_2}}\,(V_1,JdV_2)
+\sqrt{\hspace{1mm}\frac{\eta_2}{\eta_1}}\,(V_2,JdV_1)\right]
\label{bhopf}
\eeq
which in fact turns to be $\xi^{-1}\cosh\chi\,(X,JdX)$ in the 
$SU(3)\times U(1)$ limit $\eta_1 =\eta_2$, $\gamma=\sinh^2 \chi$.
Then one can rewrite the metric (\ref{dbs}) such that
$$
ds_0^2 =\frac{1}{\eta}\,\cos^2\!\mu\,ds_{FS(2)}^2 
+\frac{1}{\xi^2}\,\cos^2\!\mu
\sum_{i,\,j=1}^3 M_{ij}\,\omega^i \otimes \omega^j
+e^7 \otimes e^7.
$$
where $\omega^1 =d\mu$, $\omega^2 =d\psi$, $\omega^3 =(u,Jdu)$ and $M_{ij}$'s 
are components of the mixing matrix $M$ given by
\beq
M=\left[\begin{array}{ccc}
    f_3 & f_0 f_1 & -f_1 \\
f_0 f_1 & f_0 f_2 & -f_2 \\
   -f_1 &    -f_2 & f_2 f_0^{-1}
\end{array}\right]
\eeq
with polynomials
\bea
f_0 &=& \frac{\eta}{\sqrt{\eta_1 \eta_2}}, \nonumber\\
f_1 &=& \frac{(\eta_1-\eta_2)\cos\mu\sin\mu\cos\psi\sin\psi}
{\sqrt{\eta_1\eta_2}}, \nonumber\\
f_2 &=& \frac{(\eta_1 \cos^2\!\psi +\eta_2 \sin^2\!\psi)
\sin^2\!\mu}{\sqrt{\eta_1\eta_2}}, \nonumber\\
f_3 &=& 2\sin^2\!\mu 
+\frac{\eta}{\eta_1\eta_2}\cos^2\!\mu 
%\Bigl[\,\eta_1+\eta_2-(\eta_1-\eta_2)\cos 2\psi \,\Bigr],
\,(\eta_1 \sin^2\!\psi +\eta_2 \cos^2\!\psi) \nonumber\\
&&
 +\frac{1}{\eta}\tan^2\!\mu\sin^2\!\mu
%\Bigl[\,\eta_1+\eta_2+(\eta_1-\eta_2)\cos 2\psi \,\Bigr].
\,(\eta_1 \cos^2\!\psi +\eta_2 \sin^2\!\psi).
\eea

If the seventh frame $e^7$ we chose is indeed correct, 
the mixing matrix $M$ must be rank 2 so that its eigenvalues can be 
$\lambda_{+},\lambda_{-},0$ including zero. Equivalently, the bilinear 
form must be of the form
$$
\sum_{i,\,j=1}^3 M_{ij}\,\omega^i \otimes \omega^j = 
\lambda_{+}\,\omega^{+} \otimes \omega^{+} 
+\lambda_{-}\,\omega^{-} \otimes \omega^{-}
$$
where $\omega^{+}$, $\omega^{-}$ are eigenvectors for 
$\lambda_{+}$, $\lambda_{-}$ respectively. This is in fact the case 
as one can see by diagonalizing $M$. Two nonzero eigenvalues 
$\lambda_{+}$, $\lambda_{-}$ are solutions to the quadratic equation
$$
f(\lambda)\equiv 
\lambda^2 -\left[\,f_3 +f_2 \,(f_0+f_0^{-1})\,\right] \lambda
-(f_0+f_0^{-1})\left(f_1^2 f_0 -f_2 f_3\right)=0
$$
which says that both eigenvalues $\lambda_{+}$ and $\lambda_{-}$ are 
always positive and the corresponding eigenvectors $\omega_{+}$, $\omega_{-}$ 
are determined as
\bea
\omega^{\pm}=\frac{1}{\sqrt{(\lambda_{\pm}-\lambda_{\mp})(f_3 -\lambda_{\mp})}}
\Bigl[\,
-(f_3-\lambda_{\mp})\,\omega^1 -f_1 f_0 \,\omega^2 +f_1 \omega^3\,\Bigr]. 
\nonumber
\eea
Thus we arrive at the unwarped frames $e^{\,i}$'s given by
\bea
e^1 &=& \eta^{-{1\over2}} \cos \mu \,d\theta,\nonumber\\
e^2 &=& \eta^{-{1\over2}} \cos \mu \,{\textstyle \frac{1}{2}}
\sin\theta \,\sigma_1,\nonumber\\
e^3 &=& \eta^{-{1\over2}} \cos \mu \,{\textstyle \frac{1}{2}}
\sin\theta \,\sigma_2,\nonumber\\
e^4 &=& \eta^{-{1\over2}} \cos \mu \,{\textstyle \frac{1}{2}}
\sin\theta \cos\theta \,\sigma_3,\nonumber\\
e^5 &=& \xi^{-1}\cos\mu\, \sqrt{ \frac{\lambda_{+}}
{(\lambda_{+}-\lambda_{-})(f_3 -\lambda_{-})} }
\Bigl[\,f_1 (u,Jdu)-f_1 f_0 \,d\psi -(f_3-\lambda_{-})\,d\mu\,\Bigr],
\nonumber\\
e^6 &=& \xi^{-1}\cos\mu\, \sqrt{ \frac{\lambda_{-}}
{(\lambda_{-}-\lambda_{+})(f_3 -\lambda_{+})} }
\Bigl[\,f_1 (u,Jdu)-f_1 f_0 \,d\psi -(f_3-\lambda_{+})\,d\mu\,\Bigr],
\nonumber\\
e^7 &=& \xi^{-1} \sqrt{\mathstrut{\gamma}+1}
\left[\,\cos^2\!\mu \,(u,Jdu)+f_2 \,d\psi+f_1 \,d\mu \,\right],
\label{sieben}
\eea
where the first four frames are those on $\S^4 \cong{\bf CP}^2$ 
and $e^5$, $e^6$ are oriented to $\omega^{+}$, $\omega^{-}$ 
respectively.\footnote{
Note that the warped frames $(E^1,E^2,E^3,E^4,E^5,E^6,E^7)$ here 
are denoted by $(e^6,e^7,e^8,e^9,e^5,e^{10},e^{11})$ in Eq.\ (4.23) 
in \cite{cpw}.} 

Now we can calculate the warp factor $\Delta$ by using the local frames 
(\ref{sieben}). Some straightforward calculations show
\bea
\bigwedge_{i=1}^7 e^{\,i} = 
\sqrt{\frac{\mathstrut{\gamma}+1}{\eta^6 \eta_1 \eta_2}}\,\xi\,\Omega_7
=\frac{c^{7\over2}}{a^{9\over2} b^{9\over2} d^{5\over2}}\,\xi\,\Omega_7
\label{wedge}
\eea
where $\Omega_7$ is identified with
$$
\Omega_7 =\frac{1}{16}\sin (2\mu) \cos^4\!\mu\sin^3\!\theta\cos\theta\,
d\theta\wedge\sigma_1\wedge\sigma_2\wedge\sigma_3\wedge d\mu
\wedge d\psi\wedge (u,Jdu).
$$
This identification is correct since the prefactor in Eq.\ (\ref{wedge})
becomes 1 in the maximally symmetric $SO(8)$-invariant limit $a=b=c=d=1$.
By using Eq.\ (\ref{wedge}) in the self-consistent equation (\ref{self}), 
the warp factor is determined as
\beq
\Delta = \left(\frac{ab}{cd}\right)^{1\over6}\!c^{-1}\xi^{-{4\over3}}.
\label{su3warp}
\eeq
By using this, $\xi$ in Eq.\ (\ref{Gfull}) can be solved for 
$\Delta$ to yield
$$
G =\frac{1}{\Delta\,\xi^2 N}\,g 
%=\frac{c^{3\over2}\sqrt{\Delta}}{N}\,\left(\frac{cd}{ab}\right)^{1\over4}\!g
=\sqrt{\Delta}
\left(\frac{a^{5\over4}b^{5\over4}d^{3\over4}}{c^{3\over4}}\right)
L^2 g,
$$
which determines the warping of compact 7-dimensional space. 
Finally the warped line element $ds_7^2\equiv(dX,G\,dX)$ is determined as
\beq
ds_7^2 \equiv G_{mn}\,dy^m dy^n =\sqrt{\Delta}
\left(\frac{a^{5\over4}b^{5\over4}d^{3\over4}}{c^{3\over4}}\right)
L^2 \sum_{i=1}^7 e^{\,i} \otimes e^{\,i},
\label{su3full}
\eeq
where substitution of the local frames $e^{\,i}$ given in Eq.\ (\ref{sieben}) 
produces the warped metric $G_{mn}$ described by 
the 7-dimensional global coordinates $y^m$. 

%%%%%%%%%%%%%%%%%%%%%%%%%%%%%%%%%%%%%%%%%%%%%%%%%%%%%%%%%%%%%%%%%%%%%%%%%%%%%
\subsection{\bf The compact 7-manifold metric for various critical points 
in $SU(3)$-invariant sector}
%%%%%%%%%%%%%%%%%%%%%%%%%%%%%%%%%%%%%%%%%%%%%%%%%%%%%%%%%%%%%%%%%%%%%%%%%%%%%

There must be some comments on the local frames (\ref{sieben}) here.
First, $e^5$, $e^6$ given above are not well-defined in 
$SU(3)\times U(1)$ limit. We have $f_1=0$ in this limit so that 
the matrix $M$ shows the mixing between $\omega_1$ and $\omega_2$ only. 
Diagonalizing the matrix $M$ provides two independent frames 
$$
\rho^{-2}\xi\,d\mu \quad\mbox{and}\quad \xi^{-1} \omega \equiv
\frac{1}{2}\,\xi^{-1}\sin(2\mu)\left(\,\rho^4 (u,Jdu)-\rho^{-4}d\psi\,\right)
$$ 
as shown in \cite{cpw}. However, since either of $f_3-\lambda_{\pm}$ 
as well as $f_1$ becomes 0 in this limit, some of the coefficients 
in $e^5$, $e^6$ become indefinite if we naively take the limit. 
The careful analysis shows that 
$$
(e^5,e^6) \longrightarrow \left\{
\begin{array}{ll}
(-\rho^{-2}\xi\,d\mu,\, \xi^{-1} \omega) &
\;\mbox{for}\quad f_3 > f_2\,(f_0+f_0^{-1}) \vspace{2mm} \\
(\xi^{-1} \omega,\, +\rho^{-2}\xi\,d\mu) &
\;\mbox{for}\quad f_3 < f_2\,(f_0+f_0^{-1})
\end{array}\right..
$$ 
Because of this switching, $e^5$ and $e^6$ do not have definite physical 
meaning in $SU(3)\times U(1)$ limit although they are well-defined 
in generic $SU(3)$-invariant cases. 

Secondly, the set of local frames (\ref{sieben}) is not unique and 
any orthogonal transformation on three frames $e^5$, $e^6$ and $e^7$ 
is possible to produce the other frames 
$\tilde{e}^5$, $\tilde{e}^6$ and $\tilde{e}^7$ satisfying
$$
\tilde{e}^5 \otimes \tilde{e}^5 +\tilde{e}^6 \otimes \tilde{e}^6 
+\tilde{e}^7 \otimes \tilde{e}^7 
\equiv e^5 \otimes e^5 +e^6 \otimes e^6 +e^7 \otimes e^7. 
$$
In $G_2$ limit, we have $\eta=\eta_2=b^2$, $\eta_1=a^2$ and 
$\gamma=0$ in Eq.\ (\ref{dbs}) so that $ds_0^2$ has no contribution of 
the last term. To preserve the Fubini-Study metric on the common ${\bf CP}^2$ 
as well as the ellipsoidal deformation along $V_1$ direction, one can 
choose $\tilde{e}^5$ and $\tilde{e}^7$ so as to be fixed to $(u,Jdu)$ and 
the seventh component of $dV_1$, respectively. The remaining frame 
$\tilde{e}^6$ is determined by completing squares in $ds_0^2$. 
Then one can immediately see that $\tilde{e}^5$, $\tilde{e}^6$, 
$\tilde{e}^7$ are identified respectively with 
\beq
\frac{1}{b}\cos\mu\,(u,Jdu),\quad
\frac{\sin\psi\,d\mu+\sin\mu\cos\mu\cos\psi\,d\psi}
{b\sqrt{1-\sin^2\!\mu\cos^2\!\psi}},\quad
\frac{\xi\,d(\cos\psi\sin\mu)}{ab\sqrt{1-\sin^2\!\mu\cos^2\!\psi}}.
\label{G2base}
\eeq
However, even if we take $G_2$ limit in the local frames (\ref{sieben}), 
$e^5$, $e^6$, $e^7$ there cannot reproduce the above three frames 
without recombining into new frames via some orthogonal transformation. 
In fact, when we derived Eq.\ (\ref{sieben}) we picked up the 
broken Hopf fiber (\ref{bhopf}) as one of the local frames, expecting 
the restoration of the $U(1)$ symmetry along the Hopf fiber in 
$SU(3)\times U(1)$ limit. However in $G_2$ limit there is no restoration 
of the $U(1)$ symmetry and we have no reason to choose the broken Hopf fiber 
as one of the frames. We will discuss more on this shortly.

Let us look at the consistency of our results in both 
$SU(3)\times U(1)$ and $G_2$-invariant sectors by reconstructing 
the 7-manifold metric from the local frames (\ref{sieben}). 

$\bullet$ $SU(3)\times U(1)$-invariant sector: 
In $SU(3)\times U(1)$ limit, we have $b=1/a$ and $d=c=\cosh \chi$ 
so that Eqs.\ (\ref{su3warp}) and (\ref{su3full}) reproduce the correct 
warping of the 7-dimensional metric in section 2 \cite{cpw}: 
$$
\Delta = (\xi \cosh \chi)^{-{4\over3}},\quad 
ds_7^2 = \sqrt{\Delta}\;L^2 ds_0^2 \quad\mbox{with}\quad
\xi^2 = \rho^{-2}\cos^2\!\mu+\rho^6 \sin^2\!\mu.
$$
In this limit, we have $\eta=\rho^{-2}$, $\eta_1=\eta_2=\rho^6$ and 
$e^5$, $e^6$ become 
$$
\rho^{-2}\xi\,d\mu, \quad\mbox{and}\quad
\xi^{-1}\cos\mu\sin\mu\left(\,\rho^4 (u,Jdu)-\rho^{-4}d\psi\,\right)
$$
as mentioned before. The first four frames are those for ${\bf CP}^2$ and 
$e^7 = \xi^{-1}\cosh\chi\,(X,JdX)$ with $\chi\neq 0$ shows the stretching of 
Hopf fiber on ${\bf CP}^3$. Combining the first six frames provides
\bea
\rho^4 \sum_{i=1}^6  e^i \otimes e^i &=& \xi^2 d \mu^2 
+\cos^2\!\mu\left[\,\rho^6 ds_{FS(2)}^2 
+\xi^{-2}\sin^2\!\mu \left(\,\rho^6 (u,Jdu)-\rho^{-2}d\psi\,\right)^2
\,\right],
\nonu
\eea
which is nothing but the ellipsoidal deformation of the Fubini-Study metric 
on ${\bf CP}^3$ (See Eq.\ (\ref{FS3}) in appendix B). Therefore the compact 
7-manifold is given by the ellipsoidal deformation of stretched $\S^7$ 
as shown in \cite{cpw}.

There is an ${\cal N}=2$ supersymmetric critical point specified by 
$\rho=3^{1/8}$, $\cosh(2\chi)=2$. This is the $SU(3)\times U(1)$-invariant 
critical point found in \cite{nw}. The ${\cal N}=2$ RG flow in section 2 
carries this critical point to the $SO(8)$-invariant critical point 
specified by $\rho=1$, $\chi=0$ \cite{ap,cpw}. The $SU(4)^-$-invariant 
critical point in \cite{pw} is obtained by further taking the limit 
$\rho=1$ in $SU(3)\times U(1)$-invariant sector. Since we have 
$\xi=1$, the warp factor has no dependence on $\mu$ and is just a scaling 
factor in the sense of 7 dimensions. The compact 7-manifold is a stretched 
$\S^7$ and its ${\bf CP}^3$ base 
is described by the Fubini-Study metric given in Eq.\ (\ref{FS3}). 
The $SU(4)^-$ critical point is non-supersymmetric and there is no RG flow 
carrying it to the $SO(8)$ critical point \cite{aw}. 

$\bullet$ $G_2$-invariant sector: 
To look at this sector, it is better to use the other global coordinates 
given in appendix C. It is obtained by doing the replacement 
$$
\cos\mu = \sin\theta_6\sin\theta_7,\quad
\sin\mu\cos\psi = \cos\theta_7,\quad\mbox{and}\quad
\sin\mu\sin\psi = \cos\theta_6\sin\theta_7,
$$
with $\phi+\psi=\theta_5$ in the previous coordinates of Hopf fibration 
on ${\bf CP}^3$. The deformed norm (\ref{dnorm}) is now rewritten as
$$
\xi^2 = \eta_1\cos^2\!\theta_7
+(\eta \sin^2\!\theta_6 +\eta_2 \cos^2\!\theta_6)\sin^2\!\theta_7. 
$$
In $G_2$ limit, we have $c=a$, $d=b$ so that 
Eqs.\ (\ref{su3warp}) and (\ref{su3full}) reproduce the correct warping 
of the 7-dimensional metric in section 3 \cite{dnw,ai}:
$$
\Delta = a^{-1}\xi^{-{4\over3}},\quad 
ds_7^2 = \sqrt{\Delta\,a}\;b^2 L^2 ds_0^2, \quad\mbox{with}\quad
\xi^2 = a^2 \cos^2\!\theta_7 +b^2 \sin^2\!\theta_7.
$$
Note that $\eta=\eta_2=b^2$, $\eta_1 =a^2$ and $\gamma=0$ in this limit 
and there is no stretching of $\S^7$. 
As mentioned before, it is better to transform the last three of 
the generic frames (\ref{sieben}) into $\tilde{e}^5$, $\tilde{e}^6$, 
$\tilde{e}^7$ in Eq.\ (\ref{G2base}). They are now simply given by
$$
\tilde{e}^5 = b^{-1}\sin\theta_7\sin\theta_6\,(u,Jdu), \quad
\tilde{e}^6 = b^{-1}\sin\theta_7\,d\theta_6, \quad
\tilde{e}^7 = (ab)^{-1}\xi\,d\theta_7,
$$
and are subject to the identity
$\sum_{i=5}^7 \tilde{e}^{\,i}\otimes \tilde{e}^{\,i}\equiv
\sum_{i=5}^7 e^{\,i}\otimes e^{\,i}$.
Thus the unwarped metric for the compact 7-manifold is obtained as
$$
\sum_{i=1}^7 e^{\,i}\otimes e^{\,i}
=\frac{1}{a^2 b^2}\,\xi^2 d\theta_7^2
+\frac{1}{b^2}\sin^2\theta_7\biggl[\,d\theta_6^2 
+\sin^2\theta_6\,\left( ds_{FS(2)}^2 +(u,Jdu)^2 \right)\,\biggr],
$$
which describes the ellipsoidally deformed $\S^7$ \cite{dnw}. 
Note that the inside of the square bracket is the metric on 
$\S^6\cong G_2/SU(3)$ preserving the Fubini-Study metric on ${\bf CP}^2$. 

There is an ${\cal N}=1$ supersymmetric critical point specified by 
$a=\sqrt{\frac{6\sqrt{3}}{5}}$ and $b=\sqrt{\frac{2\sqrt{3}}{5}}$. 
This is the $G_2$-invariant critical point found in \cite{dnw}.
The ${\cal N}=1$ RG flow in section 3 carries this critical point to 
the $SO(8)$-invariant critical point specified by $a=b=1$ \cite{ai}. 
The $SO(7)^+$-invariant critical point corresponds to $a=1/b=5^{1/4}$ 
showing the ellipsoidal deformation of the 7-manifold. The difference from 
the $G_2$ critical point is that there is no field strength of 
11-dimensional gauge field in $SO(7)^+$ \cite{dn84}. 
Although this critical point is 
non-supersymmetric, there exists an RG flow connecting it to the 
$SO(8)$ critical point \cite{aw}. The $SO(7)^-$-invariant 
critical point corresponds to $a=b=\frac{\sqrt{5}}{2}$ so that 
$\xi=\frac{\sqrt{5}}{2}$ showing no deformation of $\S^7$. 
The compact 7-manifold is in fact the parallelized round $\S^7$ characterized 
by the nontrivial field strength of 11-dimensional gauge field \cite{engl}. 
The $SO(7)^-$ critical point is non-supersymmetric 
and has no RG flow carrying it to the $SO(8)$ critical point. 

%%%%%%%%%%%%%%%%%%%%%%%%%%%%%%%%%%%%%%%%%%%%%%%%%%%%%%%%%%%%%%%%%%%%%%%%%%%%%
\section{Discussions}
\setcounter{equation}{0}
%%%%%%%%%%%%%%%%%%%%%%%%%%%%%%%%%%%%%%%%%%%%%%%%%%%%%%%%%%%%%%%%%%%%%%%%%%%%%

In this section, we will discuss our obtained results (\ref{sieben}), 
(\ref{su3warp}) and (\ref{su3full}) in the point of view of 11-dimensional 
supergravity. First of all, by looking at the deformed norm $\xi$ 
in Eq.\ (\ref{dnorm}) we notice that coordinate dependence of the warp 
factor is not so simple for $G_2$-invariant sector. 
We derived the local frames (\ref{sieben}) such that the restoration of 
$U(1)$ symmetry associated with the Hopf fiber on ${\bf CP}^3$ is obvious 
in $SU(3)\times U(1)$ limit. However, such an $U(1)$ symmetry does not 
exist in $G_2$ limit and the local frames are not appropriate to look at 
the ellipsoidal deformation of the $G_2$-invariant 7-manifold. 

Therefore, the global coordinates appropriate to describe the compact 
7-manifold crucially depends on the base 6-sphere which is ${\bf CP}^3$ 
for the $SU(3)\times U(1)$-invariant sector, whereas $G_2/SU(3)$ for 
the $G_2$-invariant sector. The base manifold ${\bf CP}^3\cong\S^6$ is 
nothing but the homogeneous space $SU(4)/[SU(3)\times U(1)]$ characterized 
by the K\"{a}hler form $J$ \cite{pw}. Both $SU(4)^-$ and 
$SU(3)\times U(1)$-invariant 7-manifolds share the same $U(1)$ symmetry along 
the Hopf fiber $(X,JdX)$ so that Hopf fibration on ${\bf CP}^3$ is useful 
in those cases. On the other hand, the 7-manifolds with at least $G_2$ 
invariance, namely $SO(7)^{\pm}$ and $G_2$-invariant 7-manifolds, share 
the 6-sphere given by $G_2/SU(3)$ described as an $\S^6$ embedded in $\R^7$ 
spanned by imaginary octonions \cite{gw}. Since the ellipsoidal deformation 
is transverse to this $\S^6$, it is well described by using the 7-dimensional 
coordinates in appendix C. Therefore, the complication in the warp factor 
and in the local frames is due to the difference in $\S^6$ base between 
the two sectors. 

%In this paper we classify them into two sectors, namely $SU(3)\times U(1)$ 
%and $G_2$-invariant sectors. The compact 7-manifold has an $\S^6$ base as 
%${\bf CP}^3$ for the former, while as $G_2/SU(3)$ for the latter. 
However, in spite of the difference in the $\S^6$ base, both sectors share 
the same $\S^4\cong {\bf CP}^2$. This $\S^4$ is obvious in 
$SU(3)\times U(1)$-invariant sector, while it is implicit in 
the 6-sphere of $G_2/SU(3)$ in $G_2$-invariant sector. It was pointed out 
in \cite{cpw,jlp} that replacing the $\S^4$ with $\S^2 \times \S^2$ provides 
another 11-dimensional solution corresponding to 
the $d=4$, ${\cal N}=2$, $SU(3)\times U(1)$-invariant RG flow. 
It may be interesting to look at whether such a replacement yields 
another 11-dimensional solution corresponding to 
the $d=4$, ${\cal N}=1$, $G_2$-invariant RG flow.

As summarized in section 5, the $SU(3)$-invariant sector contains various 
critical points. It is still unclear why some of critical points have 
holographic RG flows to the $SO(8)$ critical point but is not so for others. 
To answer this question, we have to solve the 11-dimensional Einstein-Maxwell 
equations to complete the 11-dimensional lift of whole  $SU(3)$-invariant 
sector including RG flows. The 11-dimensional metric is given by 
Eq.\ (\ref{metric}) where the compact 7-manifold metric $G_{mn}$ and 
the warp factor $\Delta$ are completely determined by Eqs.\ (\ref{su3full}), 
(\ref{su3warp}) in the local frames (\ref{sieben}). The geometric parameters 
$a$, $b$, $c$, $d$ depend on the $AdS_4$ radial coordinate $r$ and are 
subject to the RG flow equations (\ref{first}) in 4-dimensional gauged 
supergravity \cite{aw}. The local frame (\ref{sieben}) found in 
this paper will be useful to achieve this work. For example, as performed 
in \cite{cpw}, one can easily make an ansatz for the 3-form gauge field 
by using the local frames. We postpone this subject for future study.
 
%%%%%%%%%%%%%%%%%%%%%%%%%%%%%%%%%%%%%%%%%%%%%%%%%%%%%%%%%%%%%%%%%%%%%%%%%%%%%
\vspace{1cm}
\centerline{\bf Acknowledgments}
 
This work by CA was supported by 
Korea Research Foundation Grant(KRF-2002-015-CS0006). 
%This research  was supported by 
%grant 2000-1-11200-001-3 from the Basic Research Program of the Korea 
%Science $\&$ Engineering Foundation. 
This research by TI was supported by 
grant 1999-2-112-003-5 from the Basic Research Program of the Korea 
Science $\&$ Engineering Foundation. 
CA is grateful to B. de Wit, H. Nicolai and N.P. Warner for 
discussions and thanks Max-Planck-Institut f\"ur Gravitationsphysik,
Albert-Einstein-Institut where part of this work was undertaken.
%\newpage
\vspace{12pt}

%%%%%%%%%%%%%%%%%%%%%%%%%%%%%%%%%%%%%%%%%%%%%%%%%%%%%%%%%%%%%%%%%%%%%%%%%%%%%
\appendix

\renewcommand{\thesection}{\large \bf \mbox{Appendix~}\Alph{section}}
\renewcommand{\theequation}{\Alph{section}\mbox{.}\arabic{equation}}

%%%%%%%%%%%%%%%%%%%%%%%%%%%%%%%%%%%%%%%%%%%%%%%%%%%%%%%%%%%%%%%%%%%%%%%%%%%%%
\section{\large \bf The 7-dimensional inverse metric encoded in 
$SU(3)$-singlet vevs of $d=4$, ${\cal N}=8$ gauged supergravity}
\setcounter{equation}{0}
%%%%%%%%%%%%%%%%%%%%%%%%%%%%%%%%%%%%%%%%%%%%%%%%%%%%%%%%%%%%%%%%%%%%%%%%%%%%%

According to the formula (\ref{7dmet}) of 7-dimensional inverse metric,
one gets all the elements of $\De^{-1} G^{AB}$ as follows.
For simplicity, we did not write them completely but some of them 
can be read off from the known expressions. For example,
$\De^{-1}G^{13}$ can be
written as $\De^{-1}G^{12}$ by replacing 
$x_2$ with $x_3$ and
$x_7$ with $ -x_6$.
We list them below. 
For simplicity, $L^{-2}\De^{-1}G^{AB}$ is denoted by $(AB)$.
\bea
(11) & = & \frac{1}{2}(ac^2 +bcd )x_2^2 + \half 
(ac^2 +bcd)x_3^2 +
\frac{1}{4}(a^3 +2a^2 b +ab^2 +ac^2-2acd +ad^2)x_4^2 \nonu \\
&& + ac^2 x_5^2 +\half(ac^2 +bcd)x_6^2+\half(ac^2+bcd)x_7^2+
\half(a^3 -ab^2-ac^2+ad^2)x_4 x_8
 \nonu \\
& & +\frac{1}{4}
(a^3 -2a^2 b +
ab^2 +ac^2+2acd +ad^2)x_8^2, \nonu \\
(12) & = & \half(-ac^2-bcd)x_1x_2 +
\half(-ac^2+bcd)x_2x_5 +\half(a^2b+ab^2-acd-bcd)x_4x_7 \nonu \\
& & +\half(a^2b-ab^2-acd+
bcd)x_7x_8, \nonu \\
(13) & = & 
\half(-ac^2-bcd)x_1x_3 +
\half(-ac^2+bcd)x_3x_5 +\half(-a^2b-ab^2+acd+bcd)x_4x_6 \nonu \\
& & +\half(-a^2b+ab^2+acd-
bcd)x_6x_8, 
\nonu \\ & = &
(12) \;\; \mbox{with} \;\; (
% \;\; \mbox{with the replacement} \;\;
x_2 \rightarrow x_3,
x_7 \rightarrow -x_6),
\nonu \\
(14) & = & -\frac{1}{4}a(a^2+2ab+b^2+c^2-2cd+d^2)x_1x_4+
\frac{1}{4}a(a^2-b^2+c^2-d^2)x_4x_5 \nonu \\
& & +\frac{1}{4}a
(a^2-b^2-c^2+d^2)x_1x_8-\frac{1}{4}a(a^2-2ab+b^2-c^2+2cd-d^2)x_5x_8, \nonu \\
(15) & = & \half(ac^2-bcd)x_2^2+(ac^2-bcd)x_3^2+\frac{1}{4}
(a^3-ab^2+ac^2-ad^2)x_4^2-ac^2x_1x_5 \nonu \\
& & +\half(ac^2-bcd)x_6^2+
\half(ac^2-bcd)x_7^2+\half(a^3+ab^2-ac^2-ad^2)x_4x_8 \nonu \\
& & +
\frac{1}{4}(a^3-ab^2+ac^2-ad^2)x_8^2, 
\nonu \\
(16) & = & 
(12) \;\; \mbox{with} \;\; ( 
%\;\; \mbox{with the replacement} \;\;
x_7 \rightarrow x_3,
x_2 \rightarrow x_6),
%\half(a^2b+ab^2-acd-bcd)x_3x_4+
%\half(-ac^2-bcd)x_1x_6+\half(-ac^2+bcd)x_5x_6 \nonu \\
%& & +\half(a^2b-ab^2-acd+
%bcd)x_3x_8, 
\nonu \\
(17) & = & 
(12) \;\; \mbox{with} \;\; (
% \;\; \mbox{with the replacement} \;\;
x_2 \rightarrow x_7,
x_7 \rightarrow -x_2),
%\half(-a^2b+ab^2+acd+bcd)x_2x_4+
%\half(-ac^2-bcd)x_1x_7+\half(-ac^2+bcd)x_5x_7 \nonu \\
%& & +\half(-a^2b+ab^2+acd-bcd)x_2x_8, 
\nonu \\
(18) & = 
& \frac{1}{4}a(-a^2+b^2+c^2-d^2)x_1x_4+\frac{1}{4}a(-a^2-2ab-b^2+
c^2+2cd+d^2)x_4x_5 \nonu \\
& & +
\frac{1}{4}a(-a^2+2ab-b^2-c^2-2cd-d^2)x_1x_8+\frac{1}{4}a(-a^2+b^2-c^2+
d^2)x_5x_8,
\nonu \\
(22) & = 
& \half(ac^2+bcd)x_1^2+bcdx_3^2 +\half(bcd+ad^2)x_4^2+\half(ac^2
+bcd)x_5^2+
(ac^2-bcd)x_1x_5 \nonu \\
& & +bcdx_6^2+
ab^2x_7^2+(-bcd+ad^2)x_4x_8+\half(bcd+ad^2)x_8^2,
\nonu \\
(23) & = & -bcdx_2x_3+(-ab^2+bcd)x_6x_7,
\nonu \\
(24) & = & 
\half(-bcd-ad^2)x_2x_4+\half(-a^2b-ab^2+acd+bcd)x_1x_7
 +\half(bcd-ad^2)x_2x_8
\nonu \\ & & +
\half(-a^2b+ab^2+acd-bcd)x_5x_7, 
\nonu \\
(25) & = & 
(12) \;\; \mbox{with} \;\; (
% \;\; \mbox{with the replacement} \;\;
x_5 \leftrightarrow x_1,
x_4 \leftrightarrow x_8),
%\half(-ac^2+bcd)x_1x_2+\half(-ac^2-bcd)x_2x_5+
%\half(a^2b-ab^2-acd+bcd)x_4x_7 \nonu \\
%& & +\half(a^2b+ab^2-acd-bcd)x_7x_8,
\nonu \\
(26) & = & 
(23) \;\; \mbox{with} \;\; (
% \;\; \mbox{with the replacement} \;\;
x_3 \rightarrow x_6,
x_6 \rightarrow -x_3),
%-bcd x_2x_6 +(ab^2-bcd)x_3x_7,
\nonu \\
(27) & = & -ab^2 x_2x_7,
\nonu \\
(28) & = & 
(24) \;\; \mbox{with} \;\; (
% \;\; \mbox{with the replacement} \;\;
x_4 \leftrightarrow x_8,
x_1 \leftrightarrow x_5),
%\half(bcd-ad^2)x_2x_4+ \half(-a^2b+ab^2+acd-bcd)x_1x_7
% + \half(-bcd-ad^2)x_2x_8
%\nonu \\ & & + 
%\half(-a^2b-ab^2+acd+bcd)x_5x_7, 
\nonu \\
(33) & = & 
(22) \;\; \mbox{with} \;\; (
% \;\; \mbox{with the replacement} \;\;
x_2 \leftrightarrow x_3,
x_7 \leftrightarrow x_6),
%\half(ac^2+bcd)x_1^2 +bcdx_2^2+\half(bcd+ad^2)x_4^2+
%(ac^2-bcd)x_1x_5 \nonu \\
%& & +
%\half(ac^2+bcd)x_5^2 + ab^2x_6^2+bcdx_7^2 +(-bcd+ad^2)x_4x_8+
%\half(bcd+ad^2)x_8^2,
\nonu \\
(34) & = & 
(24) \;\; \mbox{with} \;\; (
% \;\; \mbox{with the replacement} \;\;
x_2 \rightarrow x_3,
x_7 \rightarrow -x_6),
%\half(-bcd-ad^2)x_3x_4+ \half(a^2b+ab^2-acd-bcd)x_1x_6 
%+ \half(bcd-ad^2)x_3x_8
%\nonu \\
%& & + 
%\half(a^2b-ab^2-acd+bcd)x_5x_6, 
\nonu \\
(35) & = & 
(12) \;\; \mbox{with} \;\; (
% \;\; \mbox{with the replacement} \;\;
x_5 \leftrightarrow x_1,
x_4 \leftrightarrow x_8,
x_2 \rightarrow x_3,
x_7 \rightarrow -x_6),
%\half(-ac^2+bcd)x_1x_3+\half(-ac^2-bcd)x_3x_5+
%\half(-a^2b+ab^2+acd-bcd)x_4x_6 \nonu \\
%& & +\half(-a^2b-ab^2+acd+bcd)x_6x_8, 
\nonu \\
(36) & = & -ab^2 x_3x_6,
\nonu \\
(37) & =& 
(23) \;\; \mbox{with} \;\; (
% \;\; \mbox{with the replacement} \;\;
x_2 \rightarrow x_7,
x_7 \rightarrow -x_2),
%(ab^2-bcd) x_2x_6-bcdx_3x_7,
\nonu \\
(38) & = & 
(24) \;\; \mbox{with} \;\; (
% \;\; \mbox{with the replacement} \;\;
x_4 \leftrightarrow x_8,
x_1 \leftrightarrow x_5,
x_2 \rightarrow x_3,
x_7 \leftrightarrow -x_6),
%\half(bcd-ad^2)x_3x_4+\half(a^2b-ab^2-acd+bcd)x_1x_6
%+\half(-bcd-ad^2)x_3x_8
% \nonu \\
%& & +\half(a^2b+ab^2-acd-bcd)x_5x_6,
\nonu \\
(44) & = & 
\frac{1}{4}(a^3+2a^2b+ab^2+ac^2-2acd+ad^2)x_1^2 +
\frac{1}{2}(bcd+ad^2)x_2^2+\frac{1}{2}(bcd+ad^2)x_3^2 \nonu \\
& & +
\frac{1}{2}(a^3-ab^2+ac^2-ad^2)x_1x_5+\frac{1}{4}
(a^3-2a^2b+ab^2+ac^2+2acd+ad^2)x_5^2 \nonu \\
& & + 
\frac{1}{2}(bcd+ad^2)x_6^2 
+\frac{1}{2}(bcd+ad^2)x_7^2 +ad^2x_8^2,
\nonu \\
(45) & = & 
(18) \;\; \mbox{with} \;\; (
% \;\; \mbox{with the replacement} \;\;
x_4 \leftrightarrow x_8,
x_1 \leftrightarrow x_5),
%\frac{1}{4}a(-a^2+b^2-c^2+d^2)x_1x_4 + \frac{1}{4}a(
%-a^2+2ab-b^2-c^2-2cd-d^2)x_4x_5 \nonu \\
%& & +
% \frac{1}{4}a(-a^2-2ab-b^2+c^2+2cd+d^2)x_1x_8+ \frac{1}{4}a(
%-a^2+b^2+c^2-d^2)x_5x_8,
\nonu \\
(46) & = & 
(24) \;\; \mbox{with} \;\; (
% \;\; \mbox{with the replacement} \;\;
x_2 \rightarrow x_6,
x_7 \rightarrow x_3),
%\half(-a^2b-ab^2+acd+bcd)x_1x_3+\half(-a^2b+ab^2+acd-bcd)
%x_3x_5 \nonu \\
%& & +\half(-bcd-ad^2)x_4x_6+\half(bcd-ad^2)x_6x_8,
\nonu \\
(47) & = & 
(24) \;\; \mbox{with} \;\; (
% \;\; \mbox{with the replacement} \;\;
x_2 \rightarrow x_7,
x_7 \rightarrow -x_2),
%\half(a^2b+ab^2-acd-bcd)x_1x_2+ \half(a^2b-ab^2-acd+bcd)
%x_2x_5 \nonu \\
%& & +\half(-bcd-ad^2)x_4x_7+
%\half(bcd-ad^2)x_7x_8,
\nonu \\
(48) & = & \frac{1}{4}(a^3-ab^2-ac^2+ad^2)x_1^2 +  
\frac{1}{2}(-bcd+ad^2)x_2^2 + \frac{1}{2}(-bcd+ad^2)x_3^2 \nonu \\
& &  +
 \frac{1}{2}(a^3+ab^2-ac^2-ad^2)x_1x_5+ \frac{1}{4}(a^3-ab^2-ac^2+
ad^2)x_5^2 \nonu \\ & & 
+\half(-bcd+ad^2)x_6^2 +
 \frac{1}{2}(-bcd+ad^2)x_7^2-ad^2x_4x_8,
\nonu \\
(55) & = & 
(11) \;\; \mbox{with} \;\; (
% \;\; \mbox{with the replacement} \;\;
x_5 \leftrightarrow x_1,
x_4 \leftrightarrow x_8),
%ac^2 x_1^2 +\frac{1}{2}(ac^2+bcd)x_2^2 +\frac{1}{2}
%(ac^2+bcd)x_3^2 \nonu \\
%& & +\frac{1}{4}(a^3-2a^2b+ab^2+ac^2+2acd+ad^2)x_4^2 +
%\frac{1}{2}(ac^2+bcd)x_6^2 +\frac{1}{2}(ac^2+bcd)x_7^2
%\nonu \\ & & + 
%\frac{1}{2}(a^3-ab^2-ac^2+ad^2)x_4x_8+
%\frac{1}{4}(ab^2+ac^2-2acd+ad^2)x_8^2,
\nonu \\
(56) & = & 
(12) \;\; \mbox{with} \;\; (
% \;\; \mbox{with the replacement} \;\;
x_4 \leftrightarrow x_8,
x_1 \leftrightarrow x_5,
x_2 \leftrightarrow x_6,
x_7 \rightarrow x_3),
%\half(a^2b-ab^2-acd+bcd)x_3x_4 +\half(-ac^2+bcd)x_1x_6
%+ 
%\half(-ac^2-bcd)x_5x_6 \nonu \\ & & + 
%\half(a^2b+ab^2-acd-bcd)x_3x_8,
\nonu \\
(57) & = & 
(12) \;\; \mbox{with} \;\; (
% \;\; \mbox{with the replacement} \;\;
x_2 \rightarrow x_7,
x_5 \leftrightarrow x_1,
x_4 \leftrightarrow x_8,
x_7 \rightarrow -x_2),
%\frac{1}{2}(-a^2b+ab^2+acd-bcd)x_2x_4+
%\half(-ac^2+bcd)x_1x_7 +\half(-ac^2-bcd)x_5x_7 \nonu \\ 
%&  &+ \half(-a^2b-ab^2+acd+bcd)x_2x_8,
\nonu \\
(58) & = & 
(14) \;\; \mbox{with} \;\; (
% \;\; \mbox{with the replacement} \;\;
x_1 \leftrightarrow x_5,
x_4 \leftrightarrow x_8),
%\frac{1}{4}a(-a^2+2ab-b^2+c^2-2cd+d^2)x_1x_4+
%\frac{1}{4}a(-a^2+b^2+c^2-d^2)x_4x_5 \nonu \\
%& & +
%\frac{1}{4}a(-a^2+b^2-c^2+d^2)x_1x_8+\frac{1}{4}a(-a^2-2ab-b^2-c^2+
%2cd-d^2)x_5x_8,
\nonu \\
(66) & = & 
(22) \;\; \mbox{with} \;\; (
% \;\; \mbox{with the replacement} \;\;
x_6 \rightarrow x_2,
x_7 \leftrightarrow x_3),
%\half(ac^2+bcd)x_1^2 +bcdx_2^2 +ab^2x_3^2+\half(bcd+ad^2)x_4^2+
%(ac^2-bcd)x_1x_5 \nonu \\
%& & +\half(ac^2+bcd)x_5^2+bcdx_7^2+(-bcd+ad^2)x_4x_8+
%\half(bcd+ad^2)x_8^2,
\nonu \\
(67) & = & (-ab^2+acd)x_2x_3 -bcdx_6x_7,
\nonu \\
(68) & = & 
(24) \;\; \mbox{with} \;\; (
% \;\; \mbox{with the replacement} \;\;
x_1 \leftrightarrow x_5,
x_4 \leftrightarrow x_8,
x_2 \rightarrow x_6,
x_7 \rightarrow x_3),
%\half(-a^2b+ab^2+acd-bcd)x_1x_3+
%\half(-a^2b-ab^2+acd+bcd)x_3x_5 \nonu \\
%& & +\half(bcd-ad^2)x_4x_6+\half(-bcd-ad^2)x_6x_8,
\nonu \\
(77) & = & 
(22) \;\; \mbox{with} \;\; (
% \;\; \mbox{with the replacement} \;\;
x_7 \rightarrow x_2),
%\half(ac^2+bcd)x_1^2+ab^2x_2^2+bcdx_3^2 +\half(bcd+ad^2)x_4^2+
%(ac^2-bcd)x_1x_5 \nonu \\ & &
%+
%\half(ac^2+bcd)x_5^2+bcdx_6^2+\half(-2bcd+2ad^2+bcd)x_4x_8+\half ad^2x_8^2,
\nonu \\
(78) & = & 
(24) \;\; \mbox{with} \;\; ( 
%\;\; \mbox{with the replacement} \;\;
x_4 \leftrightarrow x_8,
x_1 \leftrightarrow x_5,
x_2 \rightarrow x_7,
x_7 \rightarrow -x_2),
%\half(a^2b-ab^2-acd+bcd)x_1x_2+
%\half(a^2b+ab^2-acd-bcd)x_2x_5 \nonu \\
%& & +\half(bcd-ad^2)x_4x_7+
%\half(-bcd-ad^2)x_7x_8,
\nonu \\
(88) & = & 
(44) \;\; \mbox{with} \;\; (
% \;\; \mbox{with the replacement} \;\;
x_5 \leftrightarrow x_1,
x_4 \leftrightarrow x_8).
%\frac{1}{4}(a^3-2a^2b+ab^2+ac^2+2acd+ad^2)x_1^2 +  
%\frac{1}{2}(bcd+ad^2)x_2^2+
% \frac{1}{2}(bcd+ad^2)x_3^2 \nonu \\
%& & + ad^2x_4^2+ 
%\frac{1}{2}(a^3-ab^2+ac^2-ad^2)x_1x_5
%+ \frac{1}{2}(bcd+ad^2)x_7^2
% \nonu \\ 
%&& +
% \frac{1}{4}(a^3+2a^2b+ab^2+ac^2-2acd+ad^2)x_5^2 + 
%\frac{1}{2}(bcd+ad^2)x_6^2
\nonu
\eea

In the $SU(4)^-$-invariant limit ($a=b=1,d=c$) the above generated results 
can be arranged into the $8 \times 8$ matrix 
$$
\Delta^{-1}G^{-1}=c^2 (I -x x^T) +(1-c^2)\,(\tilde{J}\, x)(\tilde{J}\, x)^T
$$
by introducing the K\"{a}hler form $\tilde{J}$ given by
\beq
\tilde{J} =
\left(
\begin{array}{cccccccc}
0                  & 0 & 0  & 1 & 0                  & 0 & 0 & 0 \nonu \\
0                  & 0 & 0  & 0 &                  0 & 0 & 1 & 0 \nonu \\
0                  & 0 & 0  & 0 & 0                  & -1 & 0 & 0 \nonu \\
 -1                 & 0 & 0  & 0                     & 0 & 0 & 0 & 
0 \nonu \\
0                  & 0 & 0  & 0 & 0                  & 0 & 0 & 1 \nonu \\
0                  & 0 & 1  & 0 & 0                  & 0 & 0 & 0 \nonu \\
0                  & -1 & 0  & 0 & 0                  & 0 & 0 & 0 \nonu \\
0                  & 0 & 0  & 0 & -1 & 0 & 0 & 0 
\end{array}
\right).
\label{kahler0}
\eeq
This K\"{a}hler form is transformed into the standard one $J$ given by 
Eq.\ (\ref{kahler}) in the text via $\tilde{J}=R^T J R$ with 
the orthogonal matrix
\beq
R = \left(
\begin{array}{cccccccc}
0                   & 0 & 0  & 0 &                  0 & 0 &-1 & 0 \nonu \\
0                   & 1 & 0  & 0 &                  0 & 0 & 0 & 0 \nonu \\
-\frac{1}{\sqrt{2}} & 0 & 0  & 0 & \frac{1}{\sqrt{2}} & 0 & 0 & 0 \nonu \\
0                   & 0 & 0  &-\frac{1}{\sqrt{2}} & 0 & 0 & 0 & 
\frac{1}{\sqrt{2}} \nonu \\
0                  & 0 &-1 & 0 & 0                  & 0 &  0 & 0 \nonu \\
0                  & 0 & 0 & 0 & 0                  & 1 &  0 & 0 \nonu \\
-\frac{1}{\sqrt{2}} & 0 & 0  & 0 &-\frac{1}{\sqrt{2}} & 0 &  0 & 0 \nonu \\
0                  & 0 &  0 &-\frac{1}{\sqrt{2}} & 0 & 0 &  0 & 
-\frac{1}{\sqrt{2}} \nonu
\end{array} \right).
\label{RJR}
\eeq

%%%%%%%%%%%%%%%%%%%%%%%%%%%%%%%%%%%%%%%%%%%%%%%%%%%%%%%%%%%%%%%%%%%%%%%%%%%%%
\section{\large \bf The global coordinates for $\S^7$ as 
Hopf fibration on CP$^3$}
\setcounter{equation}{0}
%%%%%%%%%%%%%%%%%%%%%%%%%%%%%%%%%%%%%%%%%%%%%%%%%%%%%%%%%%%%%%%%%%%%%%%%%%%%%

In this appendix, we summarize basic properties of Hopf fibration
and Fubini-Study metric on ${\bf CP}^n$ $(n=2,3)$ used in the text.

\vspace{12pt}

\noindent
$\bullet$ $\R^8$ embedding of Hopf fibration on ${\bf CP}^3$:
$$
X=u \cos\mu+ v \sin\mu \quad\mbox{with}\quad
u=(u^1,\dots,u^6,0,0),\quad v=(0,\dots,0,v^7,v^8).
$$
\bea
u^1+i u^2 &=& \sin\theta\cos({\textstyle {1\over2}}\alpha_1)\,
e^{\frac{i}{2}(\alpha_2+\alpha_3)}e^{i(\phi+\psi)},\nonumber\\
u^3+i u^4 &=& \sin\theta\sin({\textstyle {1\over2}}\alpha_1)\,
e^{-\frac{i}{2}(\alpha_2-\alpha_3)}e^{i(\phi+\psi)},\nonumber\\
u^5+i u^6 &=& \cos\theta\,e^{i(\phi+\psi)},\nonumber\\
v^7+i v^8 &=& e^{i\psi}.
\eea
$\bullet$ Imaginary quaternion basis:
\bea
\sigma_1 &=& \cos\alpha_3\,d\alpha_1 +\sin\alpha_1\sin\alpha_3\,d\alpha_2,
\nonumber\\
\sigma_2 &=& \sin\alpha_3\,d\alpha_1 -\sin\alpha_1\cos\alpha_3\,d\alpha_2,
\nonumber\\
\sigma_3 &=& d\alpha_3 +\cos\alpha_1\,d\alpha_2,
\eea
satisfying the Maurer-Cartan equation:
$$
d\sigma_i = \frac{1}{2}\,\epsilon_{ijk}\,\sigma_j \wedge \sigma_k.
$$
\noindent
$\bullet$ Hopf fiber on ${\bf CP}^2$:
\beq
(u,J du)=d(\phi+\psi)+\frac{1}{2}\sin^2\!\theta\,\sigma_3.
\eeq
$\bullet$ Fubini-Study metric on ${\bf CP}^2$:
\beq
ds_{FS(2)}^2 \equiv (du)^2 -(u,Jdu)^2 = 
d \theta^2 +\frac{1}{4} \sin^2\!\theta\,(\si_1^2 +\si_2^2+
\cos^2\!\theta\,\si_3^2).
\eeq
$\bullet$ Hopf fiber on ${\bf CP}^3$:
\beq
(X,JdX)=\sin^2\!\mu\,d\psi+\cos^2\!\mu\,(u,Jdu)
=d\psi+\cos^2\!\mu\left(d\phi+\frac{1}{2}\sin^2\!\theta\,\sigma_3\right).
\eeq
$\bullet$ Fubini-Study metric on ${\bf CP}^3$:
\beq
ds_{FS(3)}^2 \equiv (dX)^2 -(X,JdX)^2 =
d\mu^2+\cos^2\!\mu\left[\,ds_{FS(2)}^2 +\sin^2\!\mu
\left(d\phi+\frac{1}{2}\sin^2\!\theta\,\sigma_3\right)^2\,\right].
\label{FS3}
\eeq

\vskip 6pt

To get understanding of Hopf fibration, let us demonstrate to reconstruct 
the compact 7-manifold metric for the $SU(4)^{-}$-invariant critical point.
At this critical point, we have $a=b=1, d=c$ and $\eta=1=\eta_1=\eta_2$. 
Due to the common singularity in the $SU(3) \times U(1)$-invariant sector,  
we get either $(e^5,e^6)=
(-\,d\mu,\,  \omega)
$ or 
$(e^5,e^6)=
(\omega,\, +d\mu)
$
where 
we used the fact that $\xi =1$.
Then  $e^5 \otimes e^5+  e^6 \otimes e^6$ will lead to
\bea
d \mu^2 +\sin^2\!\mu \cos^2\!\mu 
\left( d\phi + \frac{1}{2}\sin^2\!\theta\,\si_3 \right)^2.
\nonu
\eea
Combining other $e^i, i=1,2,3,4$ with $e^5, e^6$,
the metric can be written as
\bea
\sum_{i=1}^6  e^i \otimes e^i = d \mu^2 
+\cos^2\!\mu \left[\,ds_{FS(2)}^2 +\sin^2\!\mu
\left( d\phi + \frac{1}{2}\sin^2\!\theta\,\si_3 \right)^2\,\right]
\nonu
\eea
which is nothing but the standard Fubini-Study metric on ${\bf CP}^3$ 
\cite{pw}.
Moreover, $e^7$ in this limit will be
$e^7 = c\left( \cos^2\!\mu\,(u,Jdu) +\sin^2\!\mu\,d\psi \right)$. 
Using the 
explicit form of $(u,Jdu)$ one gets 
\bea
 e^7\otimes e^7= c^2 \left[\, d\psi +\cos^2\!\mu \left( d\phi +
\frac{1}{2}\sin^2\!\theta\,\si_3  \right) \,\right]^2
\nonu
\eea
that is equal to the $U(1)$ bundle and $\psi$ is the coordinate on the
$U(1)$ fibers. According to the result of \cite{pw}, the compact 7-manifold 
metric in the above will be an Einstein which corresponds to the trivial 
$SO(8)$-invariant metric on the round 7-sphere when $c^2=1$. 
On the other hand, at the $SU(4)^{-}$-invariant critical point$(c=\sqrt{2})$,
$U(1)$ fibers over ${\bf CP}^3$ are stretched
by a factor $\sqrt{2}$ and the compact 7-manifold metric is not Einstein.

%%%%%%%%%%%%%%%%%%%%%%%%%%%%%%%%%%%%%%%%%%%%%%%%%%%%%%%%%%%%%%%%%%%%%%%%%%%%%
\section{\large \bf The global coordinates for $\S^7$ with $G_2/SU(3)$ base}
\setcounter{equation}{0}
%%%%%%%%%%%%%%%%%%%%%%%%%%%%%%%%%%%%%%%%%%%%%%%%%%%%%%%%%%%%%%%%%%%%%%%%%%%%%

In this appendix, we describe the 7-dimensional coordinatization 
appropriate for the base 6-sphere of $G_2/SU(3)$.

\vspace{12pt}

\noindent
$\bullet$ $\R^8$ embedding of $\S^7$ with $G_2/SU(3)$ base:
$$
X=u \sin\theta_6\sin\theta_7+ \tilde{v}\quad\mbox{with}\quad
u=(u^1,\dots,u^6,0,0),\quad 
\tilde{v}=(0,\dots,0,\cos\theta_7,\cos\theta_6\sin\theta_7),
$$
where $u$ is the same as above except for the 
replacement $\phi+\psi\to\theta_5$.\\

\noindent
$\bullet$ Relation to the Hopf fibration on ${\bf CP}^3$:
$$
\cos\theta_7 = \sin\mu\cos\psi,\quad 
\cos\theta_6\sin\theta_7 = \sin\mu\sin\psi,\quad
\sin\theta_6\sin\theta_7 = \cos\mu,\quad\mbox{and}\quad
\theta_5 =\phi+\psi.
$$
\beq
d\theta_7 = 
\frac{d(\cos\psi\sin\mu)}{\sqrt{1-\sin^2\!\mu\cos^2\!\psi}},\quad
\sin\theta_7\,d\theta_6 = 
\frac{\sin\psi\,d\mu+\sin\mu\cos\mu\cos\psi\,d\psi}
{\sqrt{1-\sin^2\!\mu\cos^2\!\psi}}.\label{nG2base}
\eeq
\bea
-\sin\mu\,d\mu &=& \cos\theta_6\sin\theta_7\,d\theta_6 
+\sin\theta_6\cos\theta_7\,d\theta_7, \nonumber\\
\sin^2\!\mu\,d\psi &=& \cos\theta_6\,d\theta_7 
-\sin\theta_6\cos\theta_7\sin\theta_7\,d\theta_6.
\eea
$\bullet$ The metric on $\S^6\cong G_2/SU(3)$:
\beq
d\Omega_6^2 \equiv d\theta_6^2 
+\sin^2\theta_6\,\left( ds_{FS(2)}^2 +(u,Jdu)^2 \right).
\eeq

%\newpage

%%%%%%%%%%%%%%%%%%%%%%%%%%%%%%%%%%%%%%%%%%%%%%%%%%%%%%%%%%%%%%%%%%%%%%%%%%%%%

\end{document}